\documentclass[runningheads]{llncs}
\usepackage[T1]{fontenc}
\usepackage{graphicx}
\usepackage{subcaption}
\usepackage{booktabs}
\usepackage{multirow}
\usepackage{amssymb}
\usepackage{amsmath}
\usepackage{wrapfig,booktabs}
\usepackage[font=footnotesize,labelfont=bf]{caption}
\usepackage[english]{babel}
\usepackage{hyperref}

\usepackage{color}

\begin{document}
%
\title{Discovering Sound Free-choice Workflow Nets With Non-block Structures}
%
%
\author{Tsung-Hao Huang\orcidID{0000-0002-3011-9999} \and
Wil M. P. van der Aalst\orcidID{0000-0002-0955-6940}}
\authorrunning{T. Huang and W. M. P. van der Aalst}
%
\institute{Process and Data Science (PADS), RWTH Aachen University, Aachen, Germany \\
\email{\{tsunghao.huang, wvdaalst\}@pads.rwth-aachen.de}\\
}
\maketitle              
\begin{abstract}
Process discovery aims to discover models that can explain the behaviors of event logs extracted from information systems.
While various approaches have been proposed, only a few guarantee desirable properties such as soundness and free-choice.
State-of-the-art approaches that exploit the representational bias of process trees to provide the guarantees are constrained to be block-structured.
Such constructs limit the expressive power of the discovered models, i.e., only a subset of sound free-choice workflow nets can be discovered.
To support a more flexible structural representation, we aim to discover process models that provide the same guarantees but also allow for non-block structures.
Inspired by existing works that utilize synthesis rules from the free-choice nets theory, we propose an automatic approach that incrementally adds activities to an existing process model with predefined patterns.
Playing by the rules ensures that the resulting models are always sound and free-choice.
Furthermore, the discovered models are not restricted to block structures and are thus more flexible.
The approach has been implemented in Python and tested using various real-life event logs.
The experiments show that our approach can indeed discover models with competitive quality and more flexible structures compared to the existing approach.

\end{abstract}

\section{Introduction}
Process discovery aims to construct process models that reflect the behaviors of a given event log extracted from information systems~\cite{Aalst16PMbook}.
As it is a non-trivial problem, many challenges remain.
In most cases, the one and only "best model" does not exist as there are trade-offs among the four model quality metrics, namely fitness, precision, generalization, and simplicity~\cite{Aalst16PMbook}.
In addition to the quality metrics, there exist properties that one would like to have for the discovered models.
One of the important properties is being a sound workflow net as soundness ensures the absence of deadlocks, proper completion, etc.~\cite{Aalst98worfklow} and it is a prerequisite for many crucial automated analyses such as conformance checking.
The other desirable structural property is being free-choice \cite{Aalst21usingFreeChoice}.
In free-choice nets, choices and synchronizations are separated.
This provides an easy conversion between the discovered models and many process modeling languages such as Business Process Modeling Notation (BPMN) since the equivalent constructs (dedicated split and join connectors) are naturally embedded.
Furthermore, free-choice nets have been studied extensively and thus supported by an abundance of theories~\cite{desel1995free}, which provide efficient analysis techniques.

\begin{figure}[h!]
    \centering
        \begin{subfigure}[b]{.46\linewidth}
            \includegraphics[width=\linewidth]{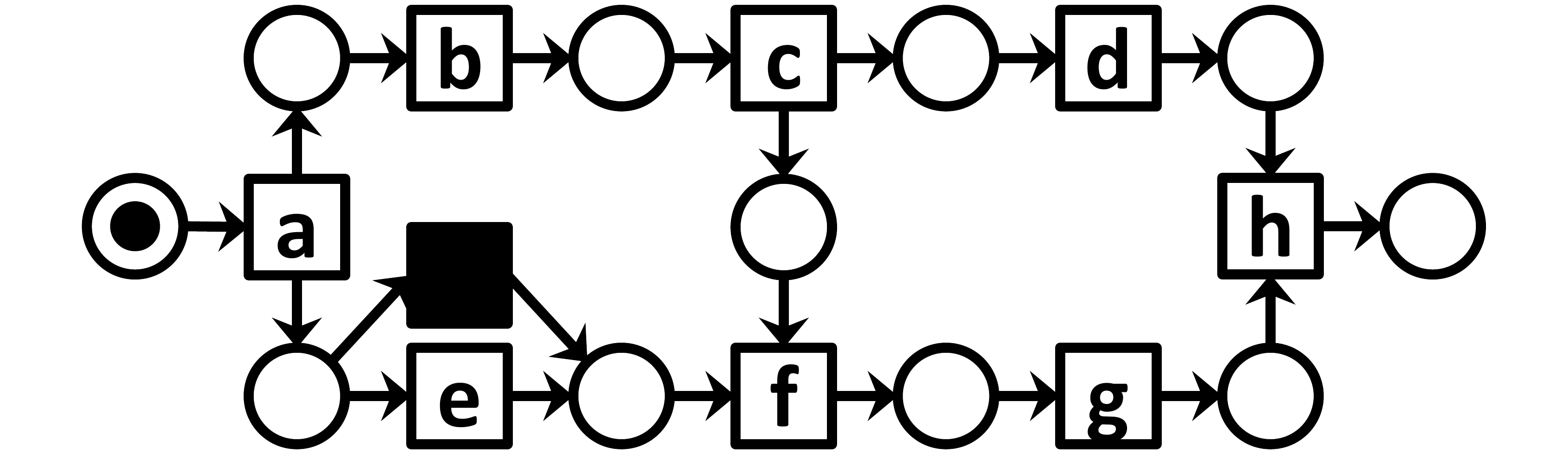}
            \caption{A model discovered by our approach. The same language cannot be expressed by the models discovered using the Inductive Miner, which uses process trees internally.}\label{fig:intro-1-1}
        \end{subfigure}
        \hspace{2em} 
        \begin{subfigure}[b]{.46\linewidth}
            \includegraphics[width=\linewidth]{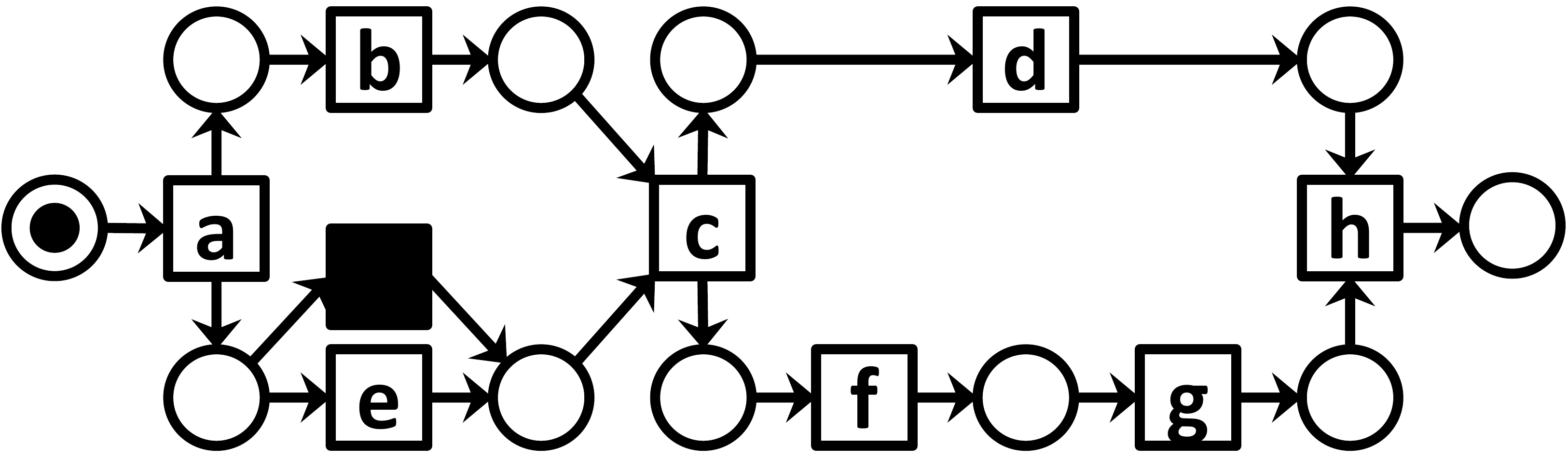}
            \caption{A model discovered by the IM using the log generated by the model in (a). The two branches before $c$ need to be synchronized first before $d$ can be executed.}\label{fig:intro-1-2}
        \end{subfigure}
    \caption{Examples showing the need for the non-block process models discovery. Note that the trace $\langle a,b,c,d,e,f,g,h \rangle$ that is possible in (a) cannot be replayed by (b).}
    \label{fig:intro}
\end{figure}

While various discovery algorithms have been proposed, only a handful of them provides such guarantees.
State-of-the-art discovery algorithms like the Inductive Miner (IM)~\cite{LeemansFA18scalableIM} are able to discover sound free-choice workflow nets by exploiting its representational bias.
However, due to the same reason, the discovered models are constrained to be block-structured.
This limits the expressive power of such models, i.e., only a subset of the sound free-choice workflow nets can be discovered.
As an example, Fig.~\ref{fig:intro-1-1} shows a sound free-choice workflow net (with non-block structures) discovered by our approach\footnote{The proposed approach has dedicated silent transitions for start and end as defined later in Def.~\ref{def:wf-net}. We dropped them here for ease of comparison.}.
The same language can never be expressed by the model discovered by IM, as shown in Fig.~\ref{fig:intro-1-2}.

In this paper, we aim to discover sound free-choice workflow nets with non-block structures.
Inspired by the interactive process discovery approach in~\cite{Dixit2019phd,DixitBA18ProDiGy}, we develop an automatic process discovery algorithm that incrementally adds activities to an existing net using synthesis rules~\cite{desel1995free}.
Since checking the feasibility for the application of the synthesis rules is computationally expensive, we use log heuristics to locate the most possible position for the to-be-added activity on the existing process model instead of evaluating all possible applications of synthesis rules as in~\cite{Dixit2019phd}.
Additionally, we identify the need for an additional rule and extend the set of patterns introduced in~\cite{DixitBA18ProDiGy}.

Playing by the rules ensures that the discovered process models by our approach are guaranteed to be sound free-choice workflow nets \cite{desel1995free,Dixit2019phd}.
Moreover, the discovered models are not constrained to block structures.
Last but not least, the level of replay fitness is guaranteed via a threshold set by the users.
The approach has been implemented in Python and evaluated using various public-available real-life event logs.
The evaluation shows that our approach is able to discover non-block structured models with competitive qualities compared to the state-of-the-art discovery algorithm.

The remainder of the paper is organized as follows. We review the related work in 
Sec.~\ref{sec:related} and introduce necessary concepts in Sec.~\ref{sec:pre}. 
Sec.~\ref{sec:approach} introduces the approach. 
Sec.~\ref{sec:experiment} presents the experiment and Sec.~\ref{sec:conclusion} concludes the paper.


\section{Related Work}\label{sec:related}
An overview of process discovery is out of the scope of this paper, we refer to \cite{AugustoCDRMMMS19PDreviewbenchmark,DongenMW09discoveryOverview} for more details.
In this section, we focus on process discovery techniques that guarantee soundness (and free-choice) properties.
Approaches like~\cite{AugustoCDRB18StructuredMiner,AugustoCDRP19splitMiner} can discover non-block structured models but cannot guarantee both properties. 
While Split Miner discovers models that are deadlock-free, they are not necessarily sound~\cite{AugustoCDRP19splitMiner}.

The family of Inductive Miner (IM) algorithms~\cite{LeemansFA18scalableIM} guarantees sound and free-choice of the discovered models by exploiting the representational bias of the process tree.
By design, a process tree represents a sound workflow net. 
It is a rooted tree where the leaf nodes are activities and the non-leaf nodes are the operators.
The hierarchical representation has a straightforward translation to Petri net.
However, the resulting models are limited to being block-structured as a process tree can only represent process models that can be separated into parts that have a single entry and exit~\cite{LeemansFA18scalableIM}.
Consequently, process trees can only represent a subset of sound workflow nets.
The same arguments hold for approaches that are based on process trees such as the Evolutionary Tree Miner (ETM)~\cite{BuijsDA12ETM} and the recently developed incremental process discovery approach~\cite{SchusterZA20Incremental}.

Applying the synthesis rules~\cite{desel1995free}, the interactive process discovery approaches developed in~\cite{DixitBA18ProDiGy,Dixit18interactive,Dixit2019phd} ensure soundness and free-choice properties.
A semi-automatic interactive tool, ProDiGy, is proposed in~\cite{DixitBA18ProDiGy} to recommend the best possible ways to add an activity to an existing model.

Our approach differs from~\cite{DixitBA18ProDiGy,Dixit18interactive,Dixit2019phd} in several ways.
First, we adopt an automatic setting as the order of adding activities is predetermined and the best modification to the existing net is selected based on the model quality.
Second, we use log heuristics to locate the most suitable position for adding the new activity instead of evaluating all the possibilities of synthesis rules applications.
Moreover, we identify the need for a new rule as the desired models often cannot be discovered without going back and forth by a combination of reduction and synthesis rules~\cite{Dixit18interactive}.
Lastly, the set of patterns is extended and formally defined.

\section{Preliminaries}\label{sec:pre}
We denote the set of all sequences over some set $A$ as $A^{*}$, the power set of $A$ as $\mathcal{P}(A)$, and the set of all multisets over $A$ as $\mathcal{B}(A)$. 
For some multiset $b \in \mathcal{B}(A)$, $b(a)$ denotes the number of times $a \in A$ appears in $b$.
For a given sequence $\sigma = \langle a_1, a_2, ..., a_n \rangle \in A^{*}$, $|\sigma| = n$ is the length of $\sigma$ and $dom(\sigma)=\{1,2,...,|\sigma|\}$ is the domain of $\sigma$. 
$\langle\rangle$ is the empty sequence.
$\sigma(i) = a_i$ denotes the i-th element of $\sigma$.
Given sequences $\sigma_1$ and $\sigma_2$, $\sigma_1 \cdot \sigma_2$ denotes the concatenation of the two.
Let $A$ be a set and $X \subseteq A$ be a subset of $A$.
For $\sigma \in A^{*}$ and $a \in A$, we define $\restriction_X \in A^{*} {\rightarrow} X^{*}$ as a projection function recursively with $\langle \rangle{\restriction_X} = \langle \rangle$, $(\langle a \rangle \cdot \sigma){\restriction_X} = \langle a \rangle \cdot \sigma{\restriction_X}$ if $a \in X$ and $(\langle a \rangle \cdot \sigma){\restriction_X} = \sigma{\restriction_X}$ if $a \notin X$.
For example, $\langle x,y,x \rangle{\restriction_{\{x,z\}}}=\langle x,x\rangle$.
Projection can also be applied to multisets of sequences, e.g., $[\langle a, b, a \rangle^{6}, \langle a, b, c \rangle^{6}, \langle b, a, c \rangle^{2}]{\restriction_{\{b,c\}}} = [\langle b\rangle^{6}, \langle b, c\rangle^{8}]$.

\begin{definition}[Trace, Log]
    A trace $\sigma \in \mathcal{U}_{A}^{*}$ is a sequence of activities, where $\mathcal{U}_{A}$ is the universe of activities.
    A log $L \in \mathcal{B}(\mathcal{U}_{A}^{*})$ is a multiset of traces.
\end{definition}

\begin{definition}[Log Properties]
    Let $L \in \mathcal{B}(\mathcal{U}_{A}^{*})$ and $a,b \in \mathcal{U}_{A}$.
    \begin{itemize}
        \item $\#(a,L) = \Sigma_{\sigma \in L}|\{i \in dom(\sigma)| \sigma(i) = a\}|$ is the times $a$ occurred in $L$.
        \item $\#(a,b,L) = \Sigma_{\sigma \in L}|\{i \in dom(\sigma){\setminus}\{|\sigma|\}| \sigma(i) = a\land \sigma(i+1) = b\}|$ is the number of direct successions from $a$ to $b$ in $L$.
        \item \begin{math}
            caus(a,b,L) = \begin{cases}
                              \frac{\#(a,b,L) - \#(b,a,L)}{\#(a,b,L) + \#(b,a,L) + 1}  & \text{if $a \neq b$}\\
                              \frac{\#(a,b,L)}{\#(a,b,L)+1} & \text{if $a = b$}
                        \end{cases}
        \end{math} is the strength of causal relation $(a,b)$.
        \item $A^{pre}_c(a,L) = \{a_{pre} \in \mathcal{U}_{A} | caus(a_{pre},a,L) \geq c\}$ is the set of $a$'s preceding activities, determined by threshold $c$.
        \item $A^{fol}_c(a,L) = \{a_{fol} \in \mathcal{U}_{A} | caus(a,a_{fol},L) \geq c\}$ is the set of $a$'s following activities, determined by threshold $c$.
        \item $A^{s}(L) = \{\sigma(1) \, |\, \sigma \in L \land \sigma \neq \langle\rangle \}$ is the set of start activities in $L$.
        \item $A^{e}(L) = \{\sigma(|\sigma|) \, |\, \sigma \in L \land \sigma \neq \langle\rangle \}$ is the set of end activities in $L$.
    \end{itemize}
\end{definition}

\begin{definition}[Petri Net, Labeled Petri Net]
    A Petri net $N = (P, T, F)$ is a tuple, where $P$ is the set of places, $T$ is the set of transitions, $P \cap T = \emptyset$, and $F\subseteq (P \times T) \cup (T \times P)$ is the set of arcs.
    A labeled Petri net $N = (P, T, F, l)$ is a Petri net $(P, T, F)$ with a labeling function $l \in T \nrightarrow \mathcal{U}_{A}$ that maps a subset of transitions to activities.
    A $t\in T$ is called invisible if $t$ is not in the domain of $l$.
\end{definition}
For any $x \in P \cup T$, $\overset{N}{\bullet} x = \{y | (y,x) \in F\}$ denotes the set of input nodes and $x \overset{N}{\bullet} = \{y | (x,y) \in F\}$ denotes the set of output nodes. 
The superscript $N$ is dropped if it is clear from the context.
The notation can be generalized to set. 
For any $X \subseteq P \cup T$, $\bullet X = \{y | \exists_{x\in X} (y,x) \in F\}$ and $X{\bullet} = \{y | \exists_{x
    \in X} (x,y) \in F\}$.

\begin{definition}[Free-choice Net]
    Let $N=(P,T,F)$ be a Petri net. $N$ is a free-choice net if for any $t_1,t_2 \in T: \bullet t_1 = \bullet t_2$ or $\bullet t_1\cap\bullet t_2 = \emptyset$.
\end{definition}
\begin{definition}[Workflow Net (WF-net)~\cite{Aalst98worfklow,Dixit2019phd}]\label{def:wf-net}
    Let $N=(P,T,F,l)$ be a labeled Petri net. $W=(P,T,F,l,p_s,p_e,\top,\bot)$ is a WF-net iff 
    (1) it has a dedicated source place $p_s \in P$: $\bullet p_s=\emptyset$ and a dedicated sink place $p_e\in P$: $p_e\bullet = \emptyset$
    (2) $\top \in T$: $\bullet\top = \{p_s\}\land p_s\bullet = \{\top\}$ and $\bot \in T$: $\bot\bullet=\{p_e\}\land \bullet p_e = \{\bot\}$ 
    (3) every node $x$ is on some path from $p_s$ to $p_e$, i.e., $\forall_{x\in P\cup T} (p_s,x)\in F^* \land (x,p_e)\in F^*$, where $F^*$ is the reflexive transitive closure of $F$.
\end{definition}

\begin{definition}[Short-circuited WF-net~\cite{Aalst98worfklow}] \label{def:sc-wf-net}
    Let $W=(P,T,F,l,p_s,p_e,\top,\bot)$ be a WF-net.
    The short-circuited WF-net of $W$, denoted by $SC(W)$, is constructed by 
    $SC(W){=}(P,T\cup\{t'\},F\cup\{(\bot,t'),(t',\top)\},l,p_s,p_e,\top,\bot)$, where $t' \notin T$.
\end{definition}

\begin{definition}[Paths, Elementary Paths] \label{def:paths}
    A path of a Petri net $N=(P,T,F)$ is a non-empty sequence of nodes $\rho = \langle x_1,x_2,...,x_n \rangle$ such that $(x_i,x_{i+1}) \in F$ for $1 \leq i < n$.
    $\rho$ is an elementary path if $x_i \neq x_j$ for $1\leq i<j\leq n$.
\end{definition}



\begin{definition} [Incidence Matrix~\cite{desel1995free}]
    Let $N = (P,T,F)$ be a Petri net. The incidence matrix $\mathbf{N}:(P \times T) \rightarrow \{-1,0,1\}$ of $N$ is defined as
    
    \begin{math}
        \mathbf{N}(p,t) =\begin{cases}
              0 & \text{if $((p, t) \notin F \land (t, p) \notin F) \lor ((p, t) \in F \land (t, p) \in F)$} \\
              -1 & \text{if $(p, t) \in F \land (t, p) \notin F$} \\
              1 & \text{if $(p, t) \notin F \land (t, p) \in F$} \\
            \end{cases}
    \end{math} \\
\end{definition}
For a Petri net $N=(P,T,F)$ and its corresponding incidence matrix $\mathbf{N}$, we use $\mathbf{N}(p)$ to denote the row vector of the corresponding $p \in P$ and $\mathbf{N}(t)$ to denote the column vector of the corresponding $t \in T$.

\begin{definition}[Linearly Dependent Nodes~\cite{desel1995free}]
    Let $N=(P,T,F)$ be a Petri net.
    $\mathbb{Q}$ is the set of rational numbers.
    A place $p$ is linearly dependent if there exists a row vector $\vec{v}:P\rightarrow\mathbb{Q}$ such that $\vec{v}(p)=0$ and $\vec{v}\cdot\mathbf{N}=\mathbf{N}(p)$. 
    A transition $t$ is linearly dependent if there exists a column vector $\vec{v}:T\rightarrow\mathbb{Q}$ such that $\vec{v}(t)=0$ and $\vec{v}\cdot\mathbf{N}=\mathbf{N}(t)$. 
\end{definition}

\begin{definition}[Synthesis Rules~\cite{desel1995free,Dixit2019phd}]
Let $W$ and $W'$ be two free-choice workflow nets, and let $SC(W) = (P,T,F,l,p_s,p_e,\top,\bot)$ and $SC(W')=(P',T',\\F',l',p_s,p_e,\top,\bot)$ be the corresponding short-circuited WF-nets:

\begin{itemize}
    \item Linear Dependent Place Rule $\psi_{P}$: $W'$ is derived from $W$ using $\psi_{P}$, i.e., $(W,W') \in \psi_{P}$ if (1) $T'=T$, $P'=P\cup\{p\}$ and $p \notin P$ is linear dependent in $SC(W')$, $F'=F \cup\widetilde{F}$ where $\widetilde{F} \subseteq ((\{p\}\times T)\cup(T\times \{p\}))$ (2) Every siphon in $SC(W')$ contains $p_s$.
    \item Linear Dependent Transition Rule $\psi_{T}$: $W'$ is derived from $W$ using $\psi_{T}$, i.e., $(W,W')\in\psi_{T}$ if $P'=P$, $T'= T\cup\{t\}$ and $t\notin T$ is linear dependent in $SC(W')$ and $F'=F \cup\widetilde{F}$ where $\widetilde{F} \subseteq ((P\times\{t\})\cup(\{t\}\times P))$, and $\forall_{t\in T\cap T'} l(t)=l'(t)$.
    \item Abstraction Rule $\psi_{A}$: $(W,W')\in\psi_{A}$ if (1) there exists a set of transitions $R \subseteq T$ and a set of places $S \subseteq P$ such that $(R\times S \subseteq F) \land (R\times S \neq \emptyset)$. (2) $SC(W')$ is constructed by adding an additional place $p\notin P$ and a transition $t\notin T$ such that $P'=P\cup\{p\}, T'=T\cup\{t\}, F'= (F{\setminus}(R\times S))\cup((R\times \{p\})\cup(\{p\}\times\{t\})\cup(\{t\}\times S))$, and $\forall_{t\in T\cap T'} l(t)=l'(t)$.
\end{itemize}
    
\end{definition}
Applying the three synthesis rules ($\psi_P,\psi_T,\psi_A$) to derive $W'$ from a sound free-choice workflow net $W$ ensures that $W'$ is also sound~\cite{Dixit18interactive,Dixit2019phd}.
Three properties need to be hold for a WF-net to be sound (1) safeness: places cannot hold multiple tokens at the same time (2) option to complete: it is always possible to reach the marking in which only the sink place is marked. (3) no dead transitions.
Next, we introduce the initial net~\cite{Dixit2019phd} and show some examples of synthesis rules applications.
\begin{figure}[t!]
    \centering
        \begin{subfigure}[b]{.112\linewidth}
            \includegraphics[width=\linewidth]{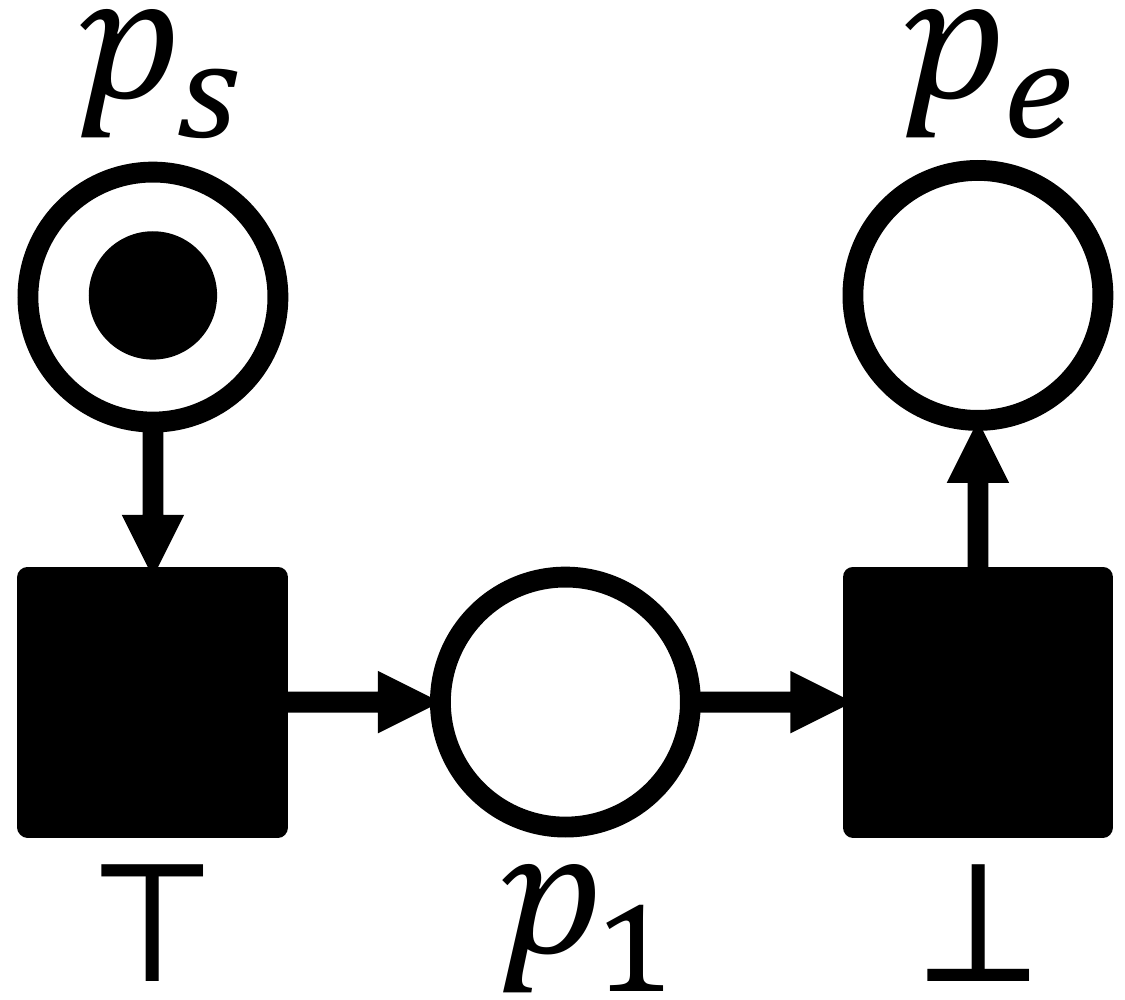}
            \caption{}\label{fig:pre-1}
        \end{subfigure}
        \hspace{0.5em} 
        \begin{subfigure}[b]{.20\linewidth}
            \includegraphics[width=\linewidth]{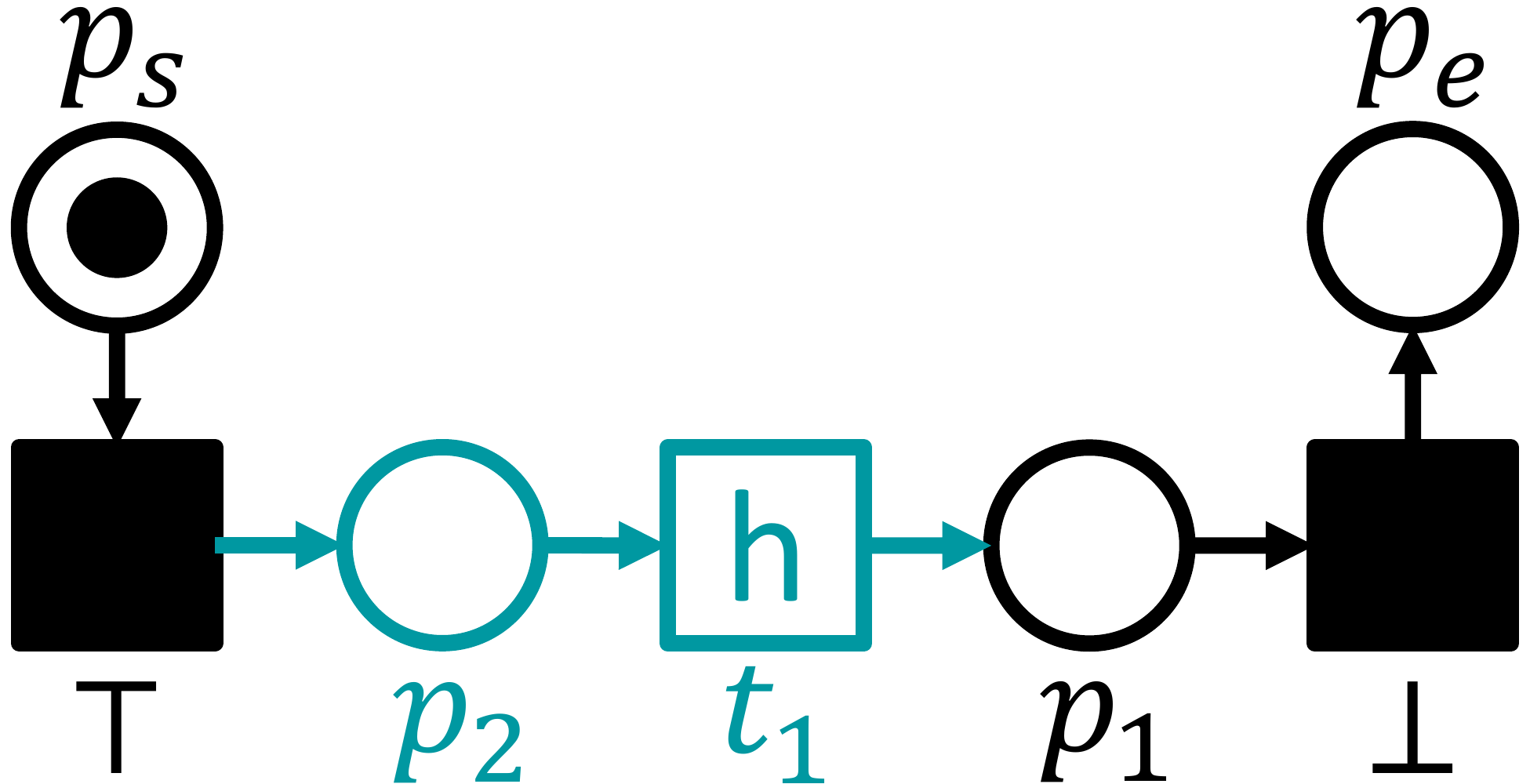}
            \caption{}\label{fig:pre-2}
        \end{subfigure}
        \hspace{0.5em} 
        \begin{subfigure}[b]{.30\linewidth}
            \includegraphics[width=\linewidth]{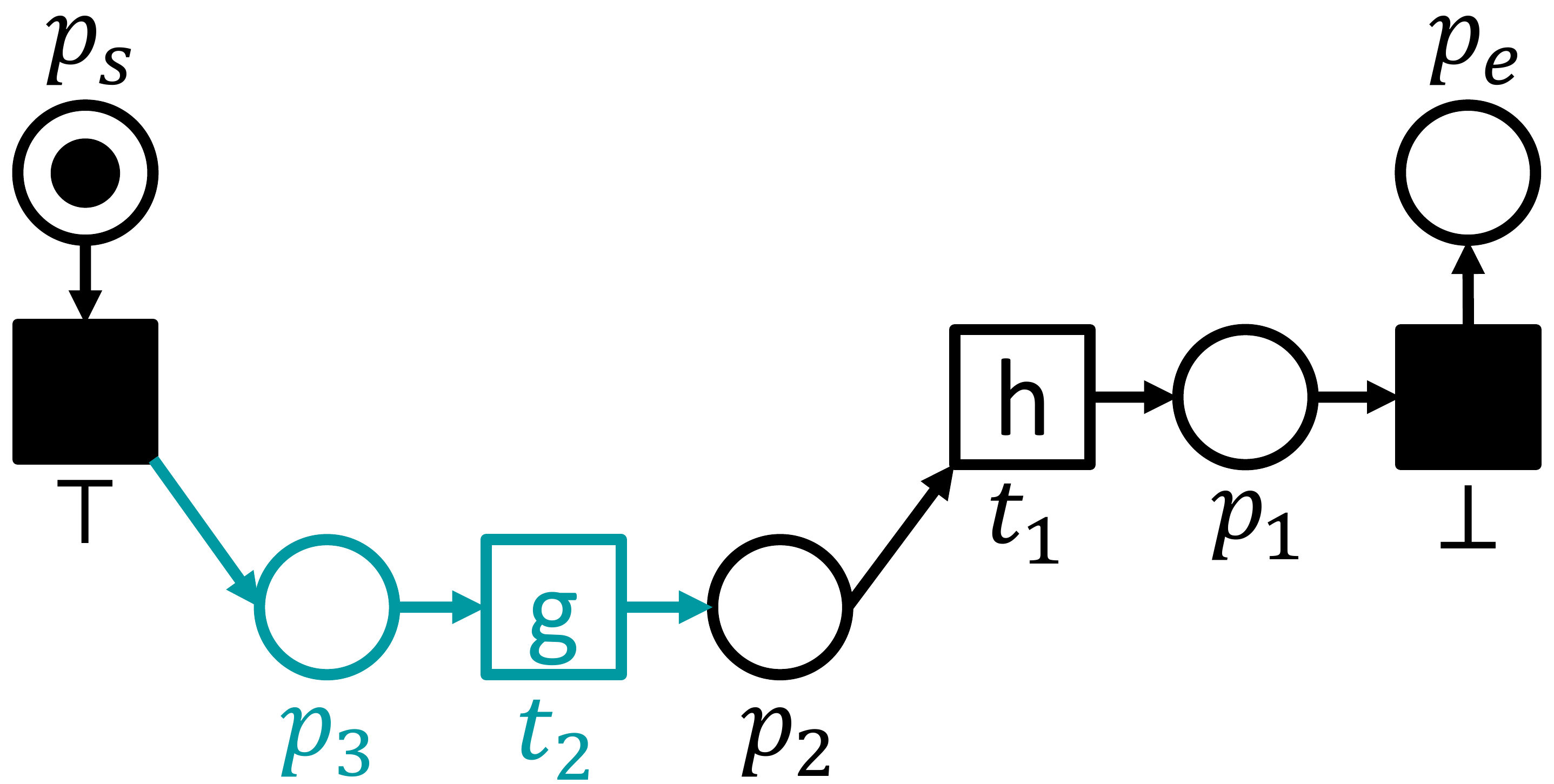}
            \caption{}\label{fig:pre-3}
        \end{subfigure}
        \hspace{0.5em} 
        \begin{subfigure}[b]{.30\linewidth}
            \includegraphics[width=\linewidth]{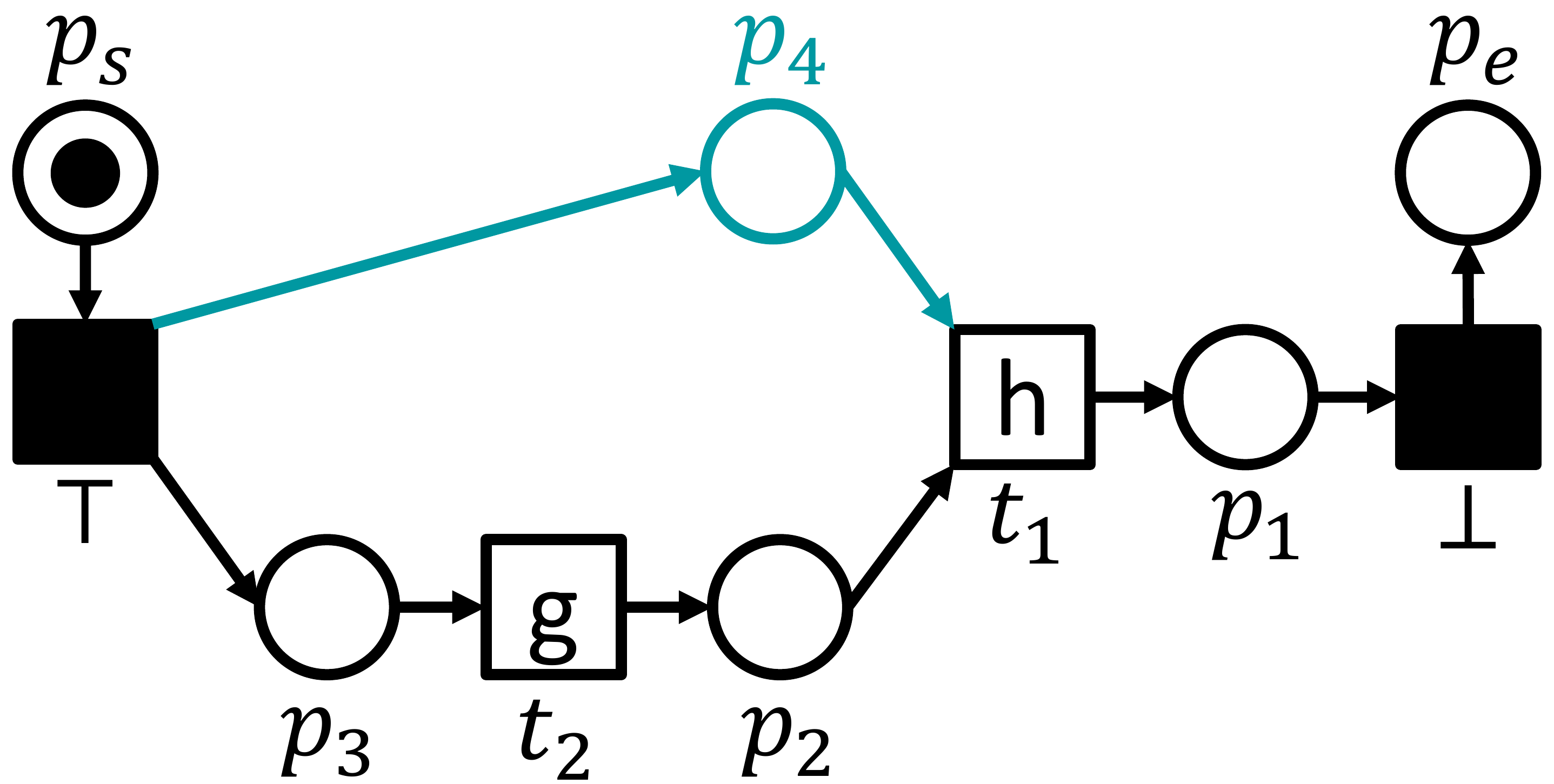}
            \caption{}\label{fig:pre-4}
        \end{subfigure}
    \caption{Examples of synthesis rules applications starting from (a) The initial net. (b) Using $\psi_A$, $p_2$ and $t_1$ are added to the initial net with $R = \{\top\}$ and $S = \{p_1\}$. (c) Using $\psi_A$, $p_3$ and $t_2$ are added to previous net with $R = \{\top\}$ and $S = \{p_2\}$. (d) $p_4$ is added using $\psi_p$ as $p_4$ is a linear combination of $p_3$ and $p_2$.}
    \label{fig:pre}
\end{figure}

\begin{definition}[Initial Net~\cite{Dixit18interactive}]\label{def:initial-net}
    Let $W=(P,T,F,l,p_s,p_e,\top,\bot)$ be a free-choice WF-net. 
    $W$ is an initial net if $P=\{p_s,p_1,p_e\}$, $T=\{\top,\bot\}$, $F=\{(p_s,\top), (\top,p_1),(p_1,\bot),(\bot,p_e)\}$.
\end{definition}
The initial net is shown in Fig.~\ref{fig:pre-1}.
Clearly, it is a sound free-choice workflow net.
Starting from the initial net, one can incrementally add additional nodes according to the synthesis rules.
Fig.~\ref{fig:pre} shows example applications of synthesis rules starting from the initial net.

\section{Approach}\label{sec:approach}
\begin{figure}[h!]
    \includegraphics[width=\textwidth]{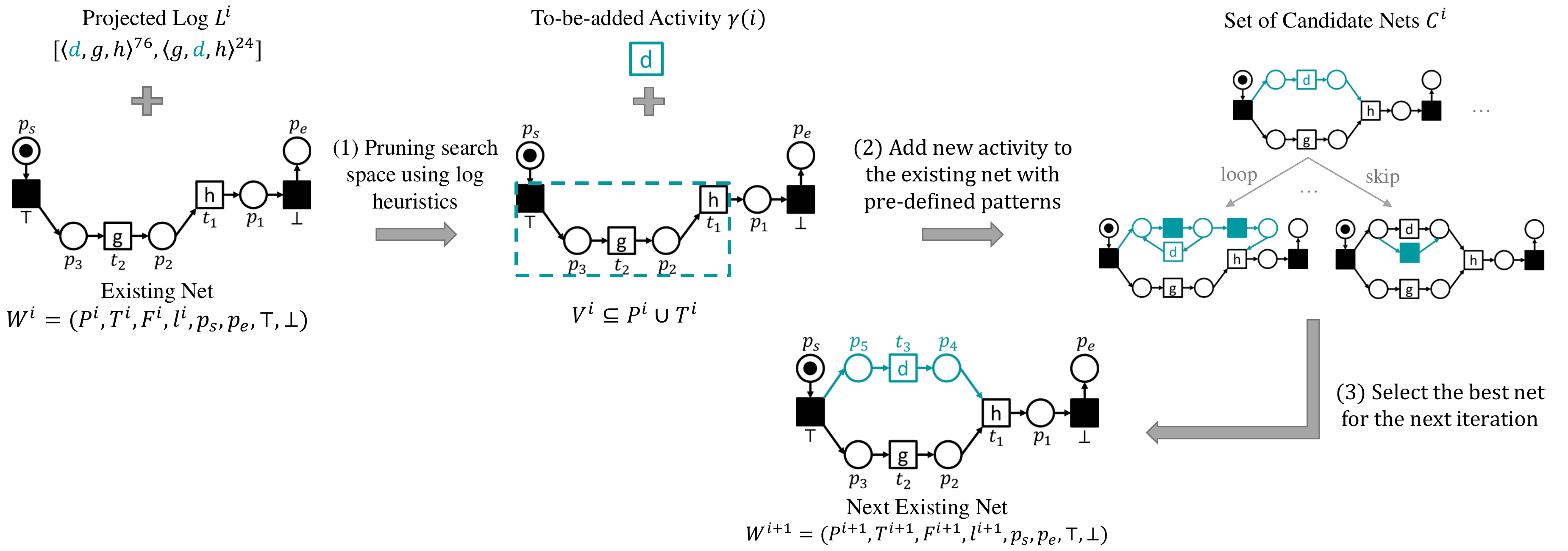}
    \caption{An example of a single iteration of our approach.} \label{fig:approach}
\end{figure}

With the necessary concepts introduced, we are now ready to introduce the approach.
We start by showing the basic idea of the approach with the help of Fig.~\ref{fig:approach} before diving into each step in detail.
Internally, the approach incrementally adds a new activity to an existing net.
The figure shows a single iteration. 
In each iteration, we have an existing model from the previous iteration and a log projected on the already added activities so far and the to-be-added one.

We start by locating the most likely position to add the new activity determined by log heuristics.
The result of this step is a subset of nodes of the existing model.
The set of nodes will then be used to prune the search space.
Then, the predefined patterns are applied to the existing net to get a set of candidate nets.
Lastly, we select the best net (next existing net) out of the candidates in terms of fitness and precision.
Note that the existing net in the first iteration is initiated by the initial net (Def.~\ref{def:initial-net}).
As a running example, consider the corresponding log that is used to discover the Petri net in Fig.\ref{fig:intro-1-1} by our approach: $L_{s} = [\langle a,b,c,d,f,g,h\rangle^{22}, \langle a,b,c,f,d,g,h\rangle^{14},\langle a,e,b,c,d,f,g,h\rangle^{13},\langle a,e,b,c,f,d,g,h\rangle^{13},\\\langle a,e,b,c,f,g,d,h\rangle^{10},\langle a,b,c,f,g,d,h\rangle^{10},\langle a,b,e,c,d,f,g,h\rangle^{6},\langle a,b,e,c,f,g,d,h\rangle^{3},\\\langle a,b,e,c,f,d,g,h\rangle^{3},\langle a,b,c,d,e,f,g,h\rangle^{2},\langle a,b,c,e,d,f,g,h\rangle^{2},\langle a,b,c,e,f,g,d,h\rangle^{1},\\\langle a,b,c,e,f,d,g,h\rangle^{1}]$.
The instances provided in Fig.~\ref{fig:approach} shows the 3rd iteration for the running example $L_s$.
In the following subsections, we introduce the details of each step.

\vspace{-1em}
\subsection{Ordering Strategies for Adding Activities} \label{subsec:ordering}
Before starting any iteration, we need to come up with an order for adding activities based on a given log $L$.
It is important as the quality of the discovered models often depends on the order of adding activities~\cite{Dixit2019phd}.
Moreover, in combination with the search space pruning, it can influence the computation time for each iteration significantly. 
In this paper, we introduce two ordering strategies.
The first one is relatively straightforward.
The activities in $L$ are simply ordered by their frequency.

\begin{definition}[Activities-Adding Order, Frequency-Based Ordering]
    Let $L \in \mathcal{B}(\mathcal{U}_{A}^{*})$ and $A=\bigcup_{\sigma\in L}\{a\in\sigma\}$.
    $\gamma\in A^{*}$ is an activities-adding order for $L$ if $\{a\in\gamma\}=A$ and $|\gamma|=|A|$.
    The frequency-based ordering is $\mathit{order_{freq}}(L) = \gamma$ such that $\gamma$ is an activities-adding order and $\forall_{1\leq i<j\leq|\gamma|} \#(\gamma(i),L) \geq \#(\gamma(j),L)$.
\end{definition}
The second ordering strategy is similar to the Breadth-first Search (BFS) algorithm. 
The advantage of this is that it also considers the closeness between activities in the log, rather than just frequency.
To explain this ordering strategy, we first define a sub-function.
\begin{definition}[Directly-Precedes Activities Sorting]
    Let $L \in \mathcal{B}(\mathcal{U}_{A}^{*})$ and $a \in \mathcal{U}_{A}$. $A {=} \{b\in\mathcal{U}_{A}|\#(b,a,L){>}0\}$ is the set of activities directly-precede $a$ in $L$ at least once and $\sigma\in A^*$.
    Directly-precedes activities sorting is $\mathit{sortPreceded}(a,L) {=} \sigma$ such that $\{b\in\sigma\}=A$ and $|\sigma|=|A|$ and $\forall_{1\leq i<j\leq |\sigma|} \,  \#(\sigma(i),a,L) \geq \#(\sigma(j),a,L)$.
\end{definition}
The function $\mathit{sortPreceded}$ takes an activity $a$ and a log $L$ to return a sequence of $a$'s directly-preceded activities $b$ that are sorted by the frequency of $\#(b,a,L)$.
Finally, we can define the BFS-based ordering strategy.
\begin{definition}[Breadth-First-Search-Based Ordering]\label{def:bfs-odering}
    Let $L \in \mathcal{B}(\mathcal{U}_{A}^{*})$ and $A{=}\bigcup_{\sigma\in L}\{a\in\sigma\}$.
    BFS-based ordering is defined as
    $\mathit{order_{BFS}}(L) {=} \gamma$, where $\gamma$ is an activities adding order for $L$ and $\gamma {=} \gamma_1 \cdot \gamma_2 \cdot ... \cdot \gamma_{|\gamma|}$, for each $1 \leq j \leq |\gamma|$,
    
    \begin{math}
        \gamma_j =\begin{cases}
              \mathit{order_{freq}}(L{\restriction_{A^e(L)}})  & \text{if $j = 1$} \\
              \mathit{sortPreceded}(\gamma(j-1), L){\restriction_{A\setminus\{\gamma(1), \gamma(2),...,\gamma(j-1)\}}} & \text{otherwise}
            \end{cases}
    \end{math}
\end{definition}
The function starts by sorting the end activities $A^e(L)$ according to their frequency in the log. 
Then, it enumerates through the sequence $\gamma$ and sorts the preceded activities of $\gamma(j-1)$ by the frequency of direct successions.
The projection function in the second case of Def.\ref{def:bfs-odering} filters out the activities that are already in $\gamma$.

Compared to the frequency-based ordering, the BFS-based ordering considers the closeness of the activities.
This allows us to add activities that are close together.
Together with the effect of the search space pruning, it is expected that BFS-based ordering would have less computation time.
Applying the function $\mathit{order_{BFS}}$ to our running example, $L_s$, we get the activities adding order as $\gamma = \mathit{order_{BFS}}(L_s)=\langle h \rangle\cdot\langle g,d \rangle\cdot\langle f \rangle\cdot\langle c,e \rangle\cdot\langle \rangle\cdot\langle b \rangle\cdot\langle a \rangle\cdot\langle \rangle = \langle h,g,d,f,c,e,b,a \rangle$.
$\gamma$ is then used to determine the order of adding activities.
Given the activities adding order $\gamma$, we define the artifacts for each iteration $i$ as followed.

\begin{definition}[Projected Log]
    Let $L \in \mathcal{B}(\mathcal{U}_{A}^{*})$ and $\gamma$ be a activities adding order for $L$.
    The projected log for $L$ in the i-th iteration is $L^{i} = L{\restriction_{\{\gamma(1),\gamma(2),...\gamma(i)\}}}$.
\end{definition}
For instance, the projected log for the running example $L_s$ for the 3rd iteration is then $L^{3}_{s}=L_s{\restriction_{\{h,g,d\}}}=[\langle d,g,h\rangle^{76}, \langle g,d,h\rangle^{24}]$, as shown in Fig.~\ref{fig:approach}.
The to-be-added activity is denoted as $\gamma(i)$, which is $\gamma(3)=d$ for the 3rd iteration.
Also, we denote the existing sound free-choice workflow net for iteration $i$ as $\mathit{W^{i}}$.
Note that for the running example, $\mathit{W^{1}}$, $\mathit{W^{2}}$, and $\mathit{W^{3}}$ are visualized in Fig.~\ref{fig:pre-1},~\ref{fig:pre-2}, and~\ref{fig:pre-3}, respectively.

\subsection{Search Space Pruning} \label{subsec:pruning}
As checking the feasibility of applying linear dependent rules $\psi_T, \psi_P$ is computationally expensive~\cite{Dixit2019phd}, it is impractical to compute all possible applications of the synthesis rules. 
Also, some of them are not of interest.
For example, as shown in Fig.~\ref{fig:approach}, it is clear that the to-be-added activity $d$ never happens after $h$ in the projected log.
Using such information, we can already eliminate the constructs (applications of synthesis rules) that allow activity $d$ to be executed after $h$.
Therefore, in each iteration $i$ we start by locating the most likely position to add $\gamma(i)$.
This helps us to restrict the application of synthesis rules on only a subset of nodes, denoted as $V^i\subseteq P^i\cup T^i$, in the existing net $W^{i}$. 
To do that, we first identify the set of preceding and following activities of $\gamma(i)$ in the projected log $L^i$, which would be $A^{pre}_c(\gamma(i),L^i)$ and $A^{fol}_c(\gamma(i),L^i)$ respectively.
Recall that $c$ is a threshold for the causal relation and can be given by users as an input.
We use $c=0.9$ as the default value.
Then, the corresponding labelled transitions are identified in $W^{i}$. 
Finally, $V^i$ is the set of all the nodes on the elementary paths from the preceding transitions to the following transitions.
If $A^{pre}_c(\gamma(i),L)$/$A^{fol}_c(\gamma(i),L)$ is an empty set, we use the $\top/\bot$ transitions.
For instance, in Fig.~\ref{fig:approach}, we identify that $A^{pre}_c(d,L_{s}^{3}){=}\emptyset$ and $A^{fol}_c(d,L_{s}^{3}){=}\{h\}$.
Therefore, we find all the nodes on the elementary paths between every node in $\{\top\}$ and every node in $\{t_1\}$.
As a result, the set $V^3 = \{\top,p_3,t_2,p_2,t_1\}$ is used to prune the search space, i.e., to constrain the application of synthesis rules.

\subsubsection{Constraining synthesis rules} 
For the abstraction rule $\psi_A$, this means that the set of transitions $R$ and the set of places $S$ used as the preconditions for applying $\psi_A$ need to be a subset of $V$, i.e., $S\subseteq V \land R\subseteq V$.
For the linear dependent rules $\psi_P$/$\psi_T$, the new place/transition ($p'$/$t'$) cannot have arcs connected to any node outside $V$.
This shortens the computation time as certain rules applications can be removed and there is no need to check their feasibility.

\subsection{Patterns} \label{subsec:patterns}
In this section, we introduce the patterns that are used to add activity $\gamma(i)$ to the existing free-choice workflow net $W^{i}$.
First, we motivate the need for an additional rule.

\subsubsection{The need for an additional rule}
\begin{figure}[h!]
    \centering
        \begin{subfigure}[b]{0.49\linewidth}
            \includegraphics[width=\linewidth]{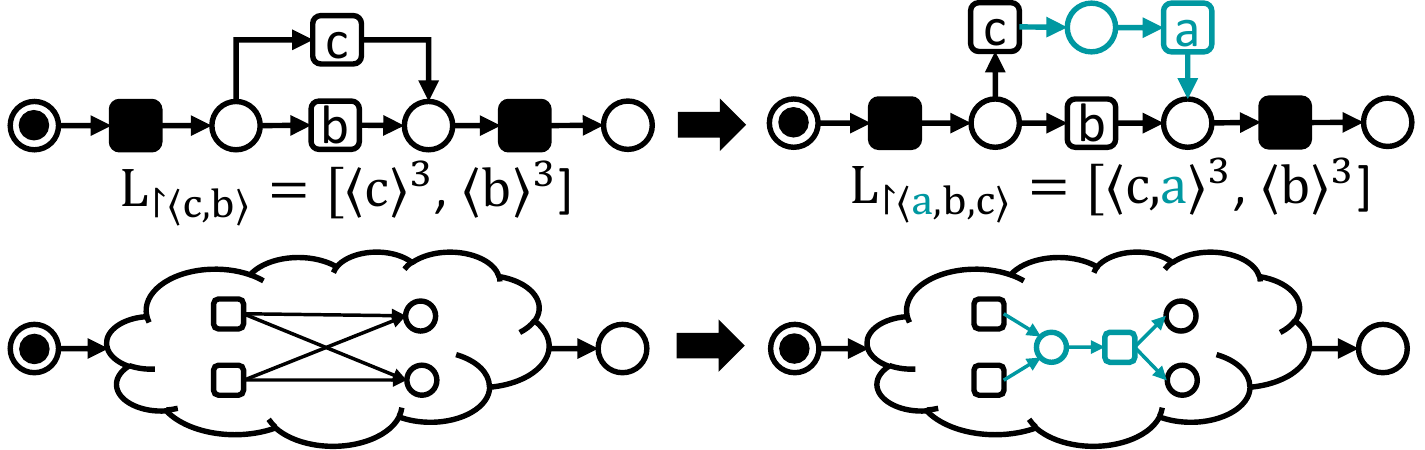}
            \caption{}\label{fig:mot-dual-abs-rule-1}
        \end{subfigure}
        \begin{subfigure}[b]{0.49\linewidth}
            \includegraphics[width=\linewidth]{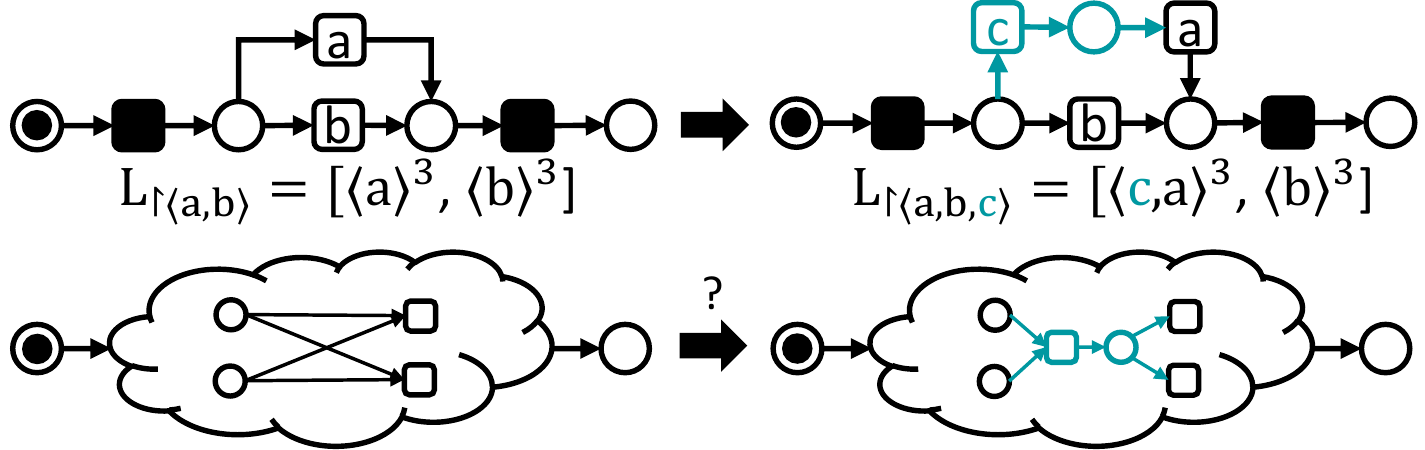}
            \caption{}\label{fig:mot-dual-abs-rule-2}
        \end{subfigure}
    \caption{Examples showing the motivation for the dual abstraction rule $\psi_D$. Although the desirable nets on the right-hand side of (a) and (b) are semantically the same, the existing synthesis rules only allow the transformation in (a). There is no rule defined for the transformation in (b).}
    \label{fig:mot-dual-abs-rule}
\end{figure}

It is proven that any sound free-choice workflow net can be constructed by the three synthesis rules $\psi_A,\psi_P,\psi_T$~\cite{desel1995free,Dixit2019phd}.
However, when applying to discover process models, the desirable model is often not possible to derive due to the existing construct.
An example is shown in Fig.~\ref{fig:mot-dual-abs-rule}.
While it is possible to add a transition labeled by $\mathit{a}$ in Fig.~\ref{fig:mot-dual-abs-rule-1}, it is not possible to derive the same net in Fig.~\ref{fig:mot-dual-abs-rule-2} as there is no rule allowing such transformation.
One possible workaround is to go back and forth by a combination of reduction and synthesis rules as suggested in \cite{Dixit18interactive}.
However, once the existing net becomes more complex, such a solution becomes infeasible to track.

We observe that in many situations (including the example in Fig.~\ref{fig:mot-dual-abs-rule-2}), the desired models cannot be constructed because there is no rule allowing one to introduce a new transition $t$ and a new place $p$ in between a set of places $S$ and a set of transitions $R$ that are fully connected, i.e., $S\times R \subseteq F$.
Therefore, we define the dual abstraction rule to allow such construct.

\begin{definition}[Dual Abstraction Rule $\psi_{D}$]\label{def:dual_abs}
    Let $W=(P,T,F,l,p_s,p_e,\top,\bot)$ and $W'=(P',T',F',l',p_s,p_e,\top,\bot)$ be two free-choice workflow nets.
    $(W,W')\in\psi_{D}$ if (1) there exists a set of places $S \subseteq P$ and a set of transitions $R \subseteq T$ such that $S \times R \subseteq F \land S\times R \neq \emptyset$. (2) $W'$ is constructed by adding an additional transition $t\notin T$ and a place $p\notin P$ such that $P' = P\cup\{p\}, T' = T\cup\{t\}, F' = (F\setminus(S \times R))\cup((S\times \{t\})\cup(\{t\}\times\{p\})\cup(\{p\}\times R))$, and $\forall_{t\in T\cap T'} l(t) = l'(t)$.
\end{definition}
As we are only interested in sound free-choice workflow nets, we need to make sure that the dual abstraction rule $\psi_{D}$ preserves soundness.
\begin{proposition}[$\psi_{D}$ preserves soundness] 
    Let $W=(P,T,F,l,p_s,p_e,\top,\bot)$, $W'=(P',T',F',l',p_s,p_e,\top,\bot)$ be free-choice workflow nets, and
    $W'$ is derived from $W$ using $\psi_{D}$, i.e., $(W,W')\in\psi_{D}$. 
    Then $W'$ is sound if $W$ is sound.
\end{proposition}

\begin{figure}[h!]
    \centering
        \begin{subfigure}[b]{.35\linewidth}
            \includegraphics[width=\linewidth]{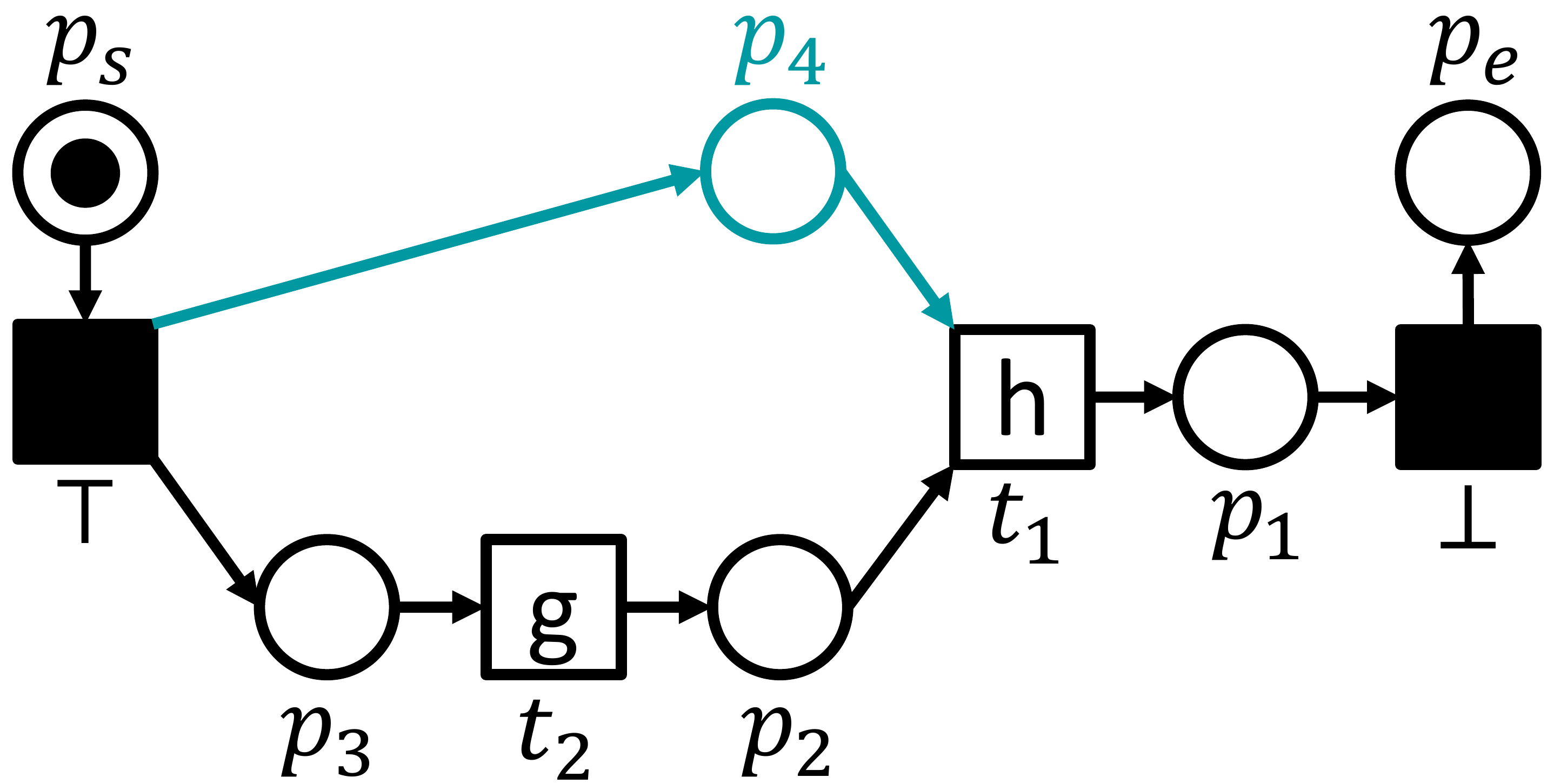}
            \caption{}\label{fig:pat_1_1_1}
        \end{subfigure}
        \hspace{2em} 
        \begin{subfigure}[b]{.35\linewidth}
            \includegraphics[width=\linewidth]{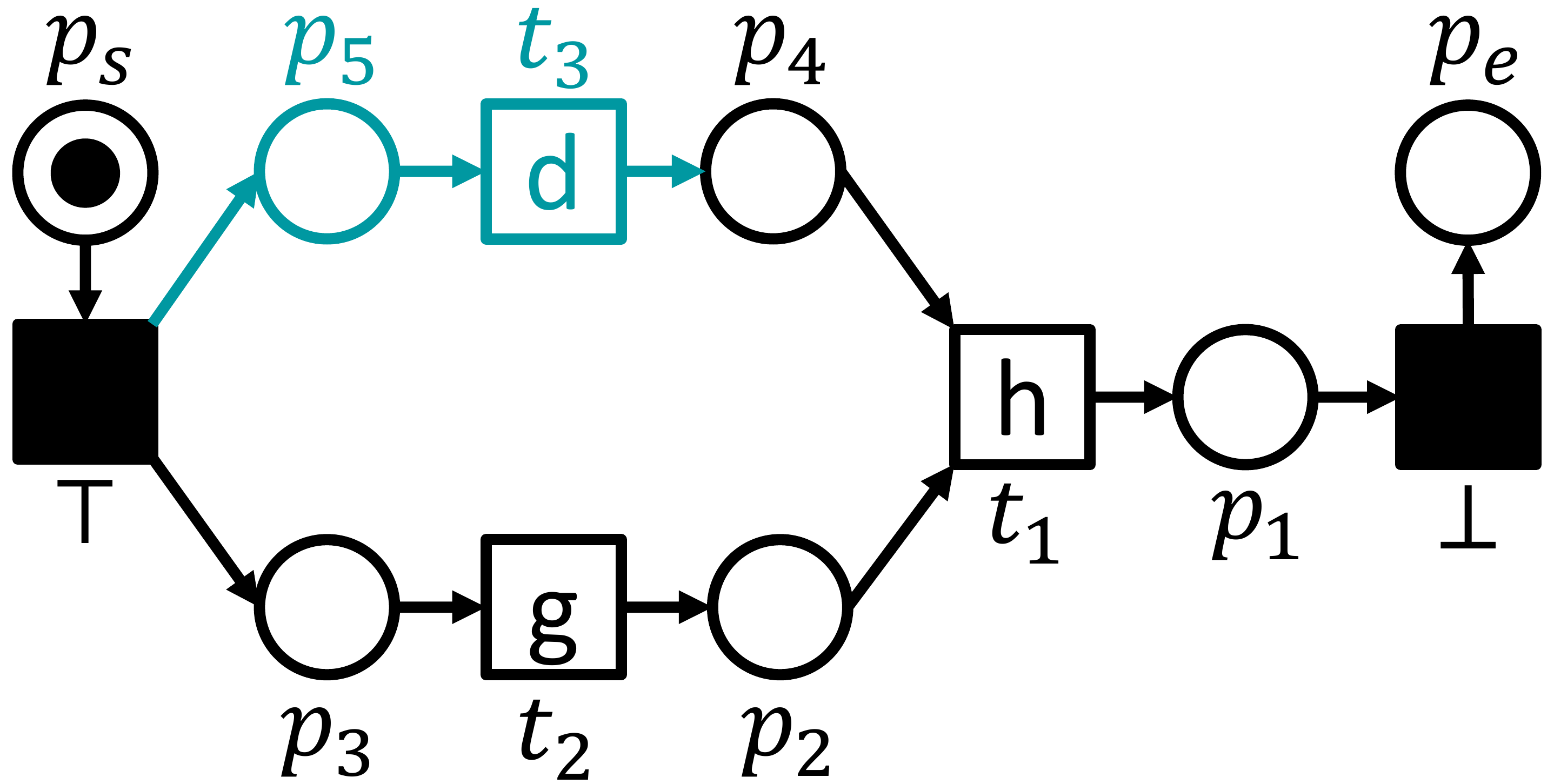}
            \caption{}\label{fig:pat_1_1_2}
        \end{subfigure}
    \caption{(a) $\psi_P$ adds a place $p_4$. (b) As every transition in $\bullet p_4$ has an arc to every place in $\{p_4\}$, one can directly apply $\psi_A$ to add $p_5$ and $t_3$.}
    \label{fig:patterns-rule-p'}
\end{figure}

\begin{proof}
    Let $t'\in T'{\setminus}T$ and $p'\in P'{\setminus}P$ be the new transition and place in $W'$. 
    Let $R=p'{\bullet}$ and $S={\bullet} t'$.
    The new net $W'$ is free-choice in only two cases. 
    Either $S={\overset{W}{\bullet}}R$ or $R=S{\overset{W}{\bullet}}$.
    In either case, any reachable marking in $(W,[p_s])$ that does not need to fire $t_R\in R$ is still reachable in $(W',[p_s])$.
    Also, the reachable markings in $(W,[p_s])$ that need to fire $t_R\in R$ can be reached in $(W',[p_s])$ as one can just add $t'$ somewhere before $t_R$ in the corresponding firing sequence.
    Then, it is trivial to see that $W'$ fulfils the three conditions of soundness if $W$ is also sound.\hfill$\square$
\end{proof}
Next, we extent the linear dependent place rule $\psi_{P}$. 
As we aim to add a transition labeled by $\gamma(i)$ to the existing labeled free-choice workflow net $W^{i}$, only adding a place $p'$ by $\psi_{P}$ does not suffice.
Hence, in our approach, an application of $\psi_{P}$ is always coupled with a directly followed application of abstraction rule $\psi_A$ to include a transition.
$\psi_A$ is applied between the added place $p'$ and its preset $\bullet p'$.
This is possible because every transition in $\bullet p'$ is connected to every place in $\{p'\}$ by definition, which satisfies the precondition of $\psi_A$.
An example is shown in Fig. \ref{fig:patterns-rule-p'}, $p_5$ and $t_3$ are added by $\psi_A$ directly after the addition of $p_4$ by $\psi_{P}$.
To be more precise, we define the extended rule, $\psi'_{P}$, that describes the pattern.
\begin{definition}[Extended Linear Dependent Place Rule $\psi'_{P}$]
    Let $W{=}(P,\\T,F,l,p_s,p_e,\top,\bot)$ and $W''{=}(P'',T'',F'',l'',p_s,p_e,\top,\bot)$ be free-choice workflow nets. 
    $(W,W''){\in} \psi'_{P}$ if (1) $\exists_{W'{=}(P',T',F',l',p_s,p_e,\top,\bot)} (W,W'){\in}\psi_{P} \land (W',W''){\in}\psi_A$ and (2) $\exists!_{p^*\in P''} (\{p^*\}{=}P'{\setminus}P) \land (((T''{\setminus}T')\times \{p^*\})\subset F'') \land ((T'\times\{p^*\})\not\subset F'')$.
\end{definition}
Then, we define the set of nets constructed by every possible single application of the rules $\psi_A, \psi'_P, \psi_T, \psi_D$.
\begin{definition}[Base Candidates Set]
    Let $W{=}(P,T,F,l,p_s,p_e,\top,\bot)$, $W'=(P',T',F',l',p_s,p_e,\top,\bot)$ be free-choice workflow nets. 
    Let $X{=}(P'{\cup} T'){\setminus}(P{\cup} T)$, $V{\subseteq}P{\cup}T$, $V'{=}(P{\cup}T){\setminus}V$, and let $a \in \mathcal{U}_A$ be an activity label.
    The base candidates set is $\mathit{base}(W,V,a){=}\{W'|((W,W')\in (\psi_{A}\cup\psi_{T}\cup\psi'_{P}\cup\psi_{D})) \land (\nexists_{x\in X} ((\{x\}\times V')\cup(V'\times \{x\}))\subseteq F') \land (l'{=}l\cup ((T'\setminus T)\times \{a\}))\}$.
\end{definition}
The base candidates set $C^{i}_{base}{=}\mathit{base}(W^i,V^i,\gamma(i))$ consists of the nets that are constructed by every possible single application of the rules $\psi_A, \psi'_P, \psi_T, \psi_D$ to add a transition labeled by $\gamma(i)$ to $W^i$ considering the the constraints on $V^{i}$.

Next, we introduce three patterns that make a transition skippable, in a strict loop, or in an optional (tau) loop.
A transition in a strict loop means that the execution of the transition is required, otherwise it is an optional loop.

\begin{figure}[h!]
    \centering
        \begin{subfigure}[b]{.35\linewidth}
            \includegraphics[width=\linewidth]{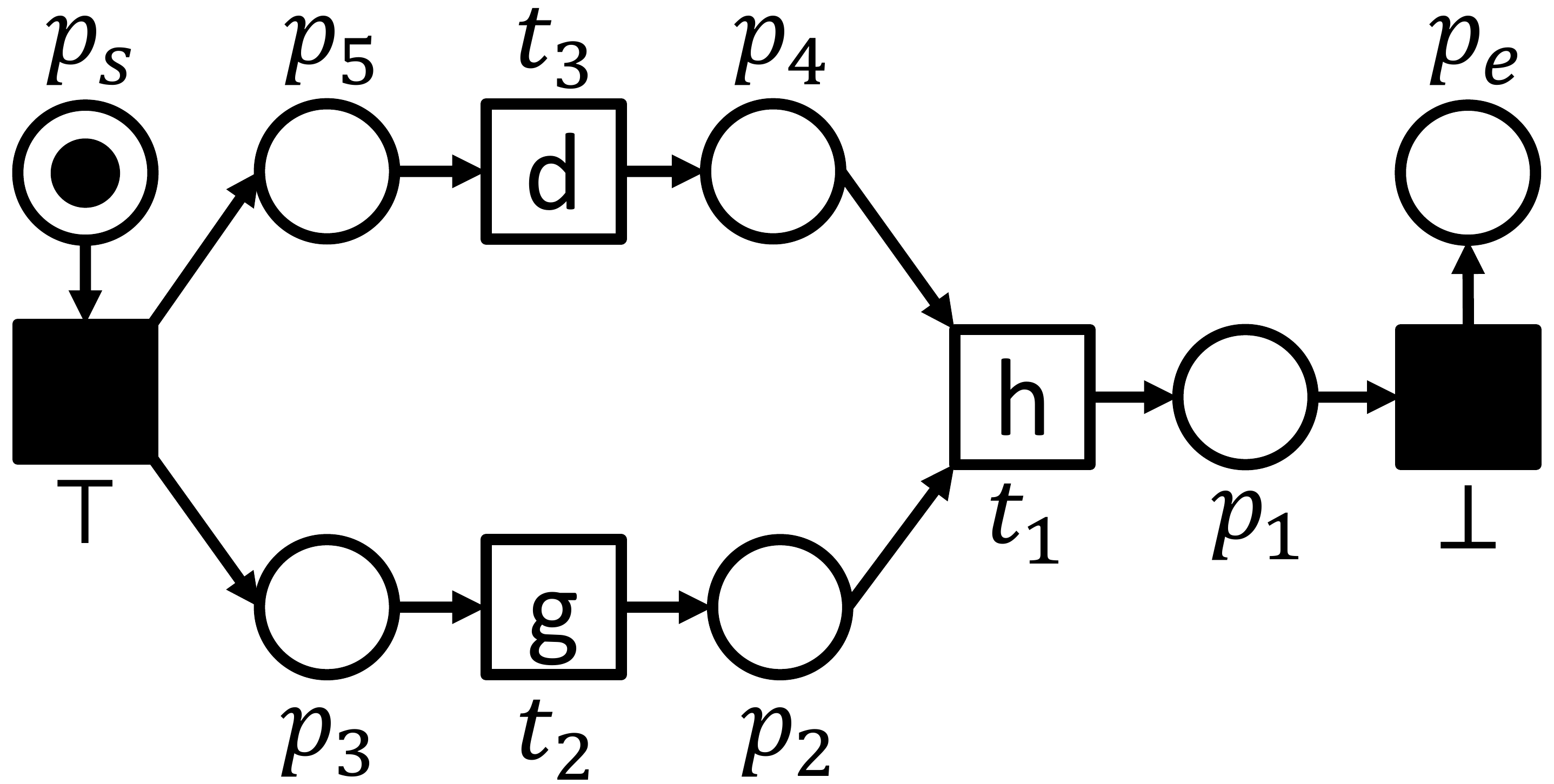}
            \caption{the net ($W$). $t_3$ (labeled by $d$) is the target transition.}\label{fig:pat_2_1_1}
        \end{subfigure}
        \hspace{2em} 
        \begin{subfigure}[b]{.35\linewidth}
            \includegraphics[width=\linewidth]{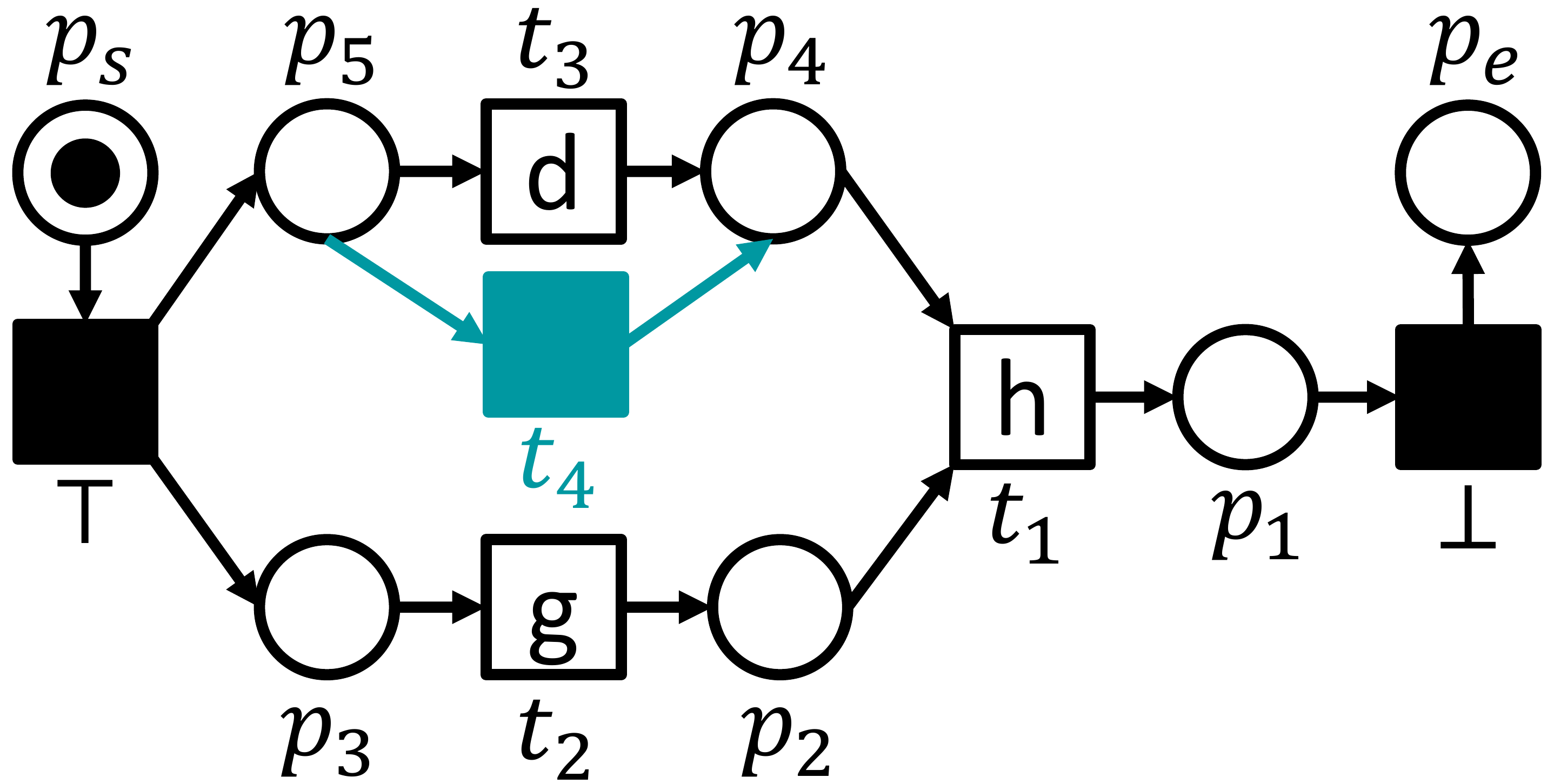}
            \caption{$skip(W,d)$ adds a silent transition $t_4$ that makes $t_3$ skippable.}\label{fig:pat_2_1_2}
        \end{subfigure}
        \par\bigskip 
        \begin{subfigure}[b]{.35\linewidth}
            \includegraphics[width=\linewidth]{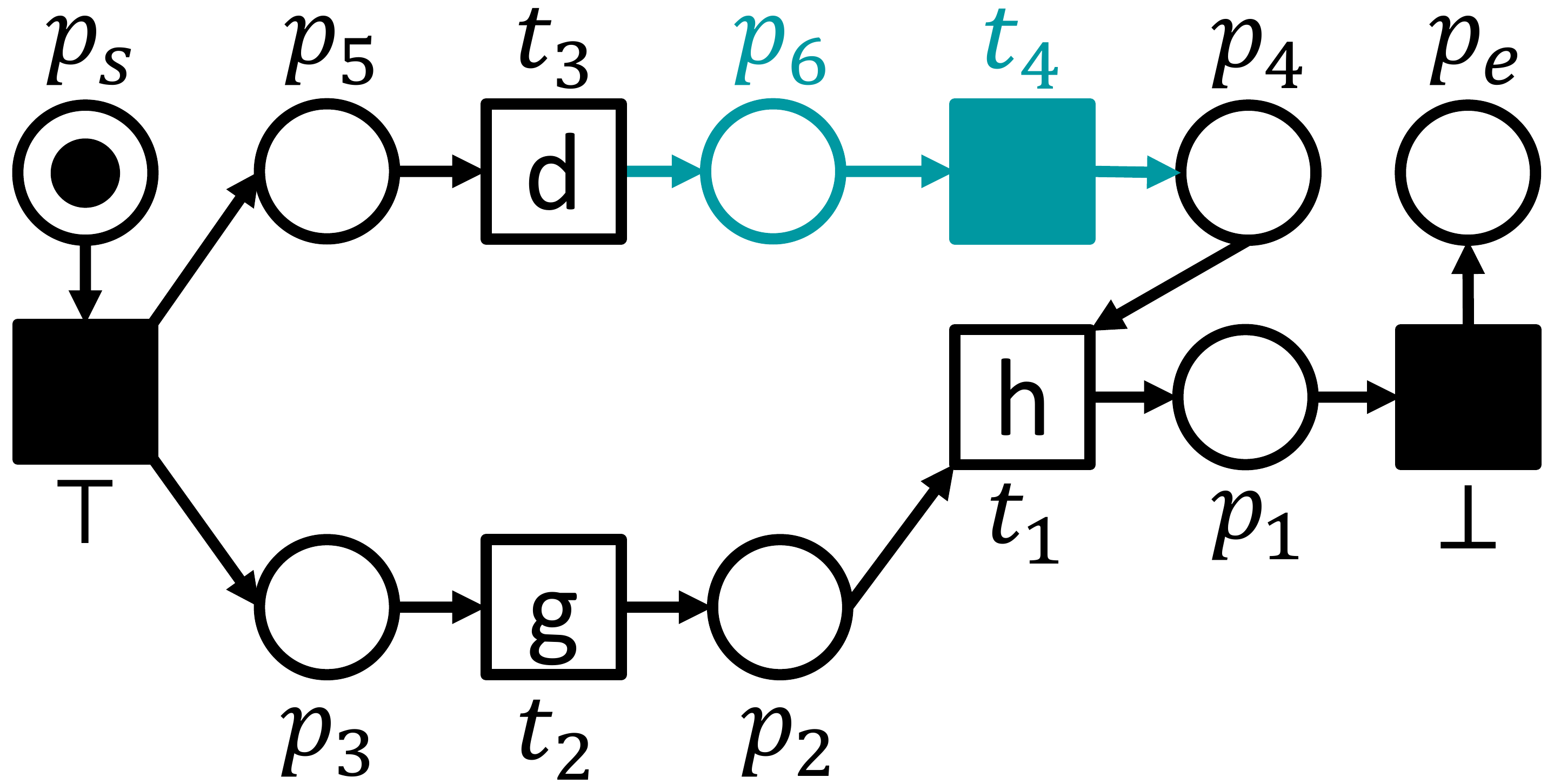}
            \caption{an intermediate net $W'$ (between (a) and (d)) constructed to keep the free-choice property.}\label{fig:pat_2_2_1}
        \end{subfigure}
        \hspace{2em} 
        \begin{subfigure}[b]{.35\linewidth}
            \includegraphics[width=\linewidth]{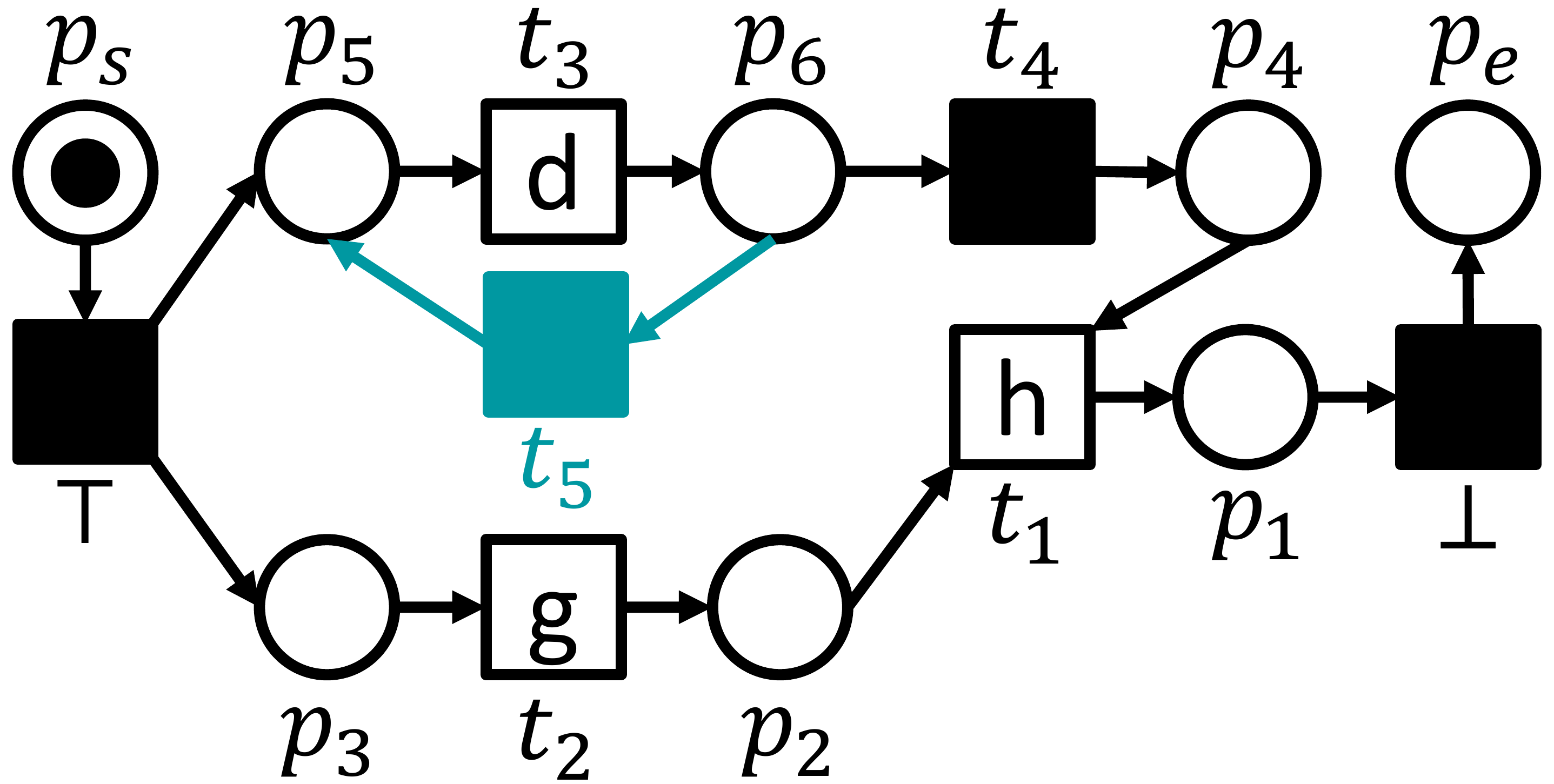}
            \caption{the resulting net of $loop_s(W,d)$, which makes $t_3$ in a loop.}\label{fig:pat_2_2_2}
        \end{subfigure}
    \caption{Examples showing how the functions are applied to derive patterns.}
    \label{fig:pat_2}
\end{figure}

\begin{definition}[Pattern-Building Functions]\label{def:pattern-building-funcs}
    Let $W{=}(P,T,F,l,p_s,p_e,\top,\bot)$ and $W'=(P',T',F',l',p_s,p_e,\top,\bot)$ be two free-choice workflow nets.
    Let $a\in \mathcal{U}_A$ be an activity label and $t_a 
    \in T: l(t_a) = a$ be the corresponding transition in $W$. We define the three pattern-building functions\footnote{The input/output nodes notations ($\bullet$) used in Def.~\ref{def:pattern-building-funcs} refer to the input net $W$. We drop the superscript for readability.} as
    \begin{itemize}
        \item $\mathit{skip}(W,a)=W'$ such that
            \begin{itemize}
                \item[--] $(W,W')\in \psi_T$
                \item[--] $F'=F\cup(\{t'\}\times t_a\bullet)\cup(\bullet t_a\times \{t'\})$ (where $t' \in T'{\setminus} T$)
                \item[--] $l'=l$ ($t'$ is a silent transition)
            \end{itemize}
        \item $\mathit{loop_s}(LW,a)$ is defined by two cases:
            \begin{enumerate}
                \item if $\nexists_{t^* \in ((t_a\bullet)\bullet)} (|{\bullet} t^*| > 1)\land({\bullet} t^*{\setminus} t_a{\bullet} \neq \emptyset)$, then $loop_s(W,a) = W'$ such that
                \begin{itemize}
                    \item[--] $(W,W')\in \psi_T$
                    \item[--] $F' = F\cup(t_a{\bullet}\times \{t'\})\cup(\{t'\}\times \bullet t_a)$ (where $t' \in T'{\setminus} T$)
                    \item[--] $l' = l$ ($t'$ is a silent transition)
                \end{itemize}       
                \item otherwise, return $\mathit{loop_s}(W', a)$ such that 
                \begin{itemize}
                    \item[--] $(W,W')\in \psi_A$
                    \item[--] $((\{t_a\}\times(P'{\setminus} P))\in F')\land((\{t_a\}\times P) \notin F')$ 
                    \item[--] $l'=l$
                \end{itemize}
            \end{enumerate}
        \item $\mathit{loop}_{\tau}(W,a)$ is defined by two cases:
            \begin{enumerate}
                \item if $\nexists_{t^* \in ((t_a\bullet)\bullet)} (|{\bullet} t^*| > 1)\land({\bullet} t^*{\setminus} t_a{\bullet}\neq\emptyset)$, then $\mathit{loop}_{\tau}(W,a){=}W'$ such that
                \begin{itemize}
                    \item[--] $(W,W')\in \psi_T$
                    \item[--] $F' = F\cup(t_a{\bullet}\times \{t'\})\cup(\{t'\}\times \bullet t_a)$ (where $t' \in T'{\setminus} T$)
                    \item[--] $l' = (l{\setminus}\{(t_a, a)\})\cup\{(t',a)\}$ (the labels of $t_a$ and $t'$ are swapped)
                \end{itemize}       
                \item otherwise, return $\mathit{loop}_{\tau}(W', a)$ such that 
                \begin{itemize}
                    \item[--] $(W,W')\in \psi_A$
                    \item[--] $((\{t_a\}\times(P'{\setminus} P))\in F')\land((\{t_a\}\times P) \notin F')$
                    \item[--] $l'=l$
                \end{itemize}
            \end{enumerate}
    \end{itemize} 
\end{definition}
The second case of the loop functions is there to keep the free-choice property.
To illustrate the ideas using the running example, consider the net shown in Fig.~\ref{fig:pat_2_1_1} as the input net $W$ and $t_3$ (labeled by $d$) is the transition for which we are going to apply the functions to derive patterns.
Fig.~\ref{fig:pat_2_1_2} shows that function $\mathit{skip}(W,d)$ simply adds a silent transition $t_4$ with the same connection as $t_3$ to $W$.
Fig.~\ref{fig:pat_2_2_1} and~\ref{fig:pat_2_2_2} show an application of $\mathit{loop_s}(W,d)$ and illustrate the need for the two cases for the loop functions.
As shown in Fig.~\ref{fig:pat_2_2_1}, the second case of $\mathit{loop_s}$ is applied since there exists a transition $t_1 \in ((t_3\bullet)\bullet)$ with more than one place in its preset ($|\bullet t_1| > 1$) and $\bullet t_1{\setminus} t_3\bullet\neq\emptyset$.
Therefore, $W'$ (Fig.~\ref{fig:pat_2_2_1}) is first constructed by adding $p_6$ and $t_4$. 
Then, the function returns $\mathit{loop_s}(W',d)$.
Now, the first case should be applied.
In this case, $t_5$ is added with the reverse connections of $t_3$.
As indicated, the second case in the loop functions helps to keep the free-choice property.
Imagine a net that is constructed by adding $t'$ to the net in Fig.~\ref{fig:pat_2_1_1} with connections $(p_4,t')$ and $(t',p_5)$.
Such a net makes $t_3$ in a loop but it is no longer a free-choice net.
The constructs of $\mathit{loop_s}$ and $\mathit{loop}_{\tau}$ are almost the same, the difference is that the labels of $t_3$ and the silent transition $t_5$ are swapped.

Finally, to get the set of candidate nets $C^{i}$, we apply the three pattern-building functions to every net $W \in C^{i}_{base}$.
Observe that all the nets in Fig.~\ref{fig:pat_2} are elements of $C^3$.

\subsection{Selection and Fall-through}

\subsubsection{Selection}
In the last step, we select the next existing net $W^{i+1}$ from the set of candidates $C^i$ evaluated by the projected log $L^{i}$. 
The selection is done in a stepwise manner. 
We first try to filter out the candidates that do not reach a user-defined replay fitness threshold $\theta$ and then select the best net out of the rest in terms of F1 score, which is calculated as the harmonic mean of fitness and precision.
We use alignment-based fitness \cite{AalstAD12Alignment} and precision \cite{AdriansyahMCDA15Precision}.
\subsubsection{Fall-through}
If none of the nets in $C^{i}$ reach the fitness threshold $\theta$, we adopt a fall-through.
This is done by going back to Step 2, where $\gamma(i)$ is added to $W^{i}=(P^{i},T^{i},F^{i},l^{i},p_s,p_e,\top,\bot)$, but without the constraints of $V^i$. This can also be seen as setting $V^i=P^{i}\cup T^{i}$.
In this case, a new place $p'$ with arcs $\{(\top,p'),(p',\bot)\}$ can be always added by $\psi_P$ as $p$ is linear dependent on $p_s$ and $p_e$.
Then, the patterns building functions can be applied to ensure that the fitness threshold $\theta$ is guaranteed in every iteration.

\vspace{-1em}
\section{Evaluation}\label{sec:experiment}
\vspace{-1em}
In this section, we present the experiments conducted to evaluate our approach. 
The presented approach in this paper is implemented in Python using PM4Py\footnote{https://pm4py.fit.fraunhofer.de/} and can be accessed here\footnote{https://github.com/tsunghao-huang/synthesisRulesMiner}.
As mentioned, the algorithm takes as inputs a log and three parameters including two thresholds $\theta, c$, and the types of ordering strategy.
Using this implementation, we conduct three experiments to address the following questions 
(1) How effective are the pre-defined patterns? 
(2) What are the effects of the ordering strategy on the model quality and the execution time?
(3) Can the model quality be improved by the non-block structures?

\vspace{-1em}
\subsection{Experiment Setup}
\vspace{-0.5em}
\textbf{Dataset:}
We use four public available real-life event logs, which are BPI2017\footnote{https://doi.org/10.4121/uuid:3926db30-f712-4394-aebc-75976070e91f},
helpdesk\footnote{https://doi.org/10.4121/uuid:0c60edf1-6f83-4e75-9367-4c63b3e9d5bb}, hospitalBilling\footnote{https://doi.org/10.4121/uuid:76c46b83-c930-4798-a1c9-4be94dfeb741}, 
and traffic\footnote{https://doi.org/10.4121/uuid:270fd440-1057-4fb9-89a9-b699b47990f5} respectively.
BPI2017 is split into two sub logs, BPI2017A and BPI2017O, using the event prefixes.
To focus on the mainstream behaviors, the logs are filtered to include at least 95\% of the cases. 


\vspace{-1em}
\subsubsection{Experiment 1 (Effectiveness of patterns):}
The first experiment aims to evaluate how effective are the pre-defined patterns. 
As our approach is based on \cite{Dixit2019phd}, this can be evaluated by comparing the quality of the intermediate models of our approach to the ones from ProDiGy \cite{DixitBA18ProDiGy}, which adopts a similar setting.
To conduct the experiment, we follow the top recommendation of ProDiGy in every step to get the intermediate models and compare the models' quality with ours.
We use the projected log of every iteration to evaluate the model obtained after adding additional activity to the model.
To have a fair comparison, we force our approach to use the same order of adding activities from ProDiGy. 

\vspace{-1em}
\subsubsection{Experiment 2 (Effects of Ordering Strategy \& Search Space Pruning):}
The order of adding activities to the log is crucial to our approach as model quality is highly dependent on the order~\cite{Dixit2019phd}.
Moreover, the order can influence the execution time due to its influence on the search space pruning.
Therefore, we would like to investigate the effects of the ordering strategy on the model quality and the execution time.
To set up the experiment, we apply the approach to the five event logs using the two different ordering strategies while keeping the other two parameters at the same values.
We evaluate the model quality in terms of fitness, precision, and F1 score. In addition, we keep track of the ratio of the reduced nodes, which is calculated by $\frac{|V^i|}{|P^{i}\cup T^{i}|}$. 
This gives us an indication of the effectiveness of search space pruning.

\vspace{-1em}
\subsubsection{Experiment 3 (Effects of non-block structures):}
In this experiment, we compare our approach to the state-of-the-art: Inductive Miner - Infrequent (IMf) \cite{LeemansFA18scalableIM}.
As the models discovered by IMf are guaranteed to be sound free-choice workflow net as well, comparing our approach with IMf enables us to see if the models can benefit from the non-block structures discovered by our approach.
For each event log, we apply IMf using five different values ($[0.1, 0.2, 0.3, 0.4, 0.5]$) for the filter threshold and choose the best model (by F1 score) to compare the quality with the ones discovered by our approach in experiment 2.

For all the experiments, we use the alignment-based approaches to calculate fitness \cite{AalstAD12Alignment} and precision \cite{AdriansyahMCDA15Precision}.
We also calculate the F1 score as the harmonic mean of the fitness and precision.

\subsection{Results}

\subsubsection{Effectiveness of Patterns}
\begin{figure}[h!]
    \centering
    \vspace{-2em}
        \begin{subfigure}[b]{0.48\linewidth}
            \includegraphics[width=\linewidth]{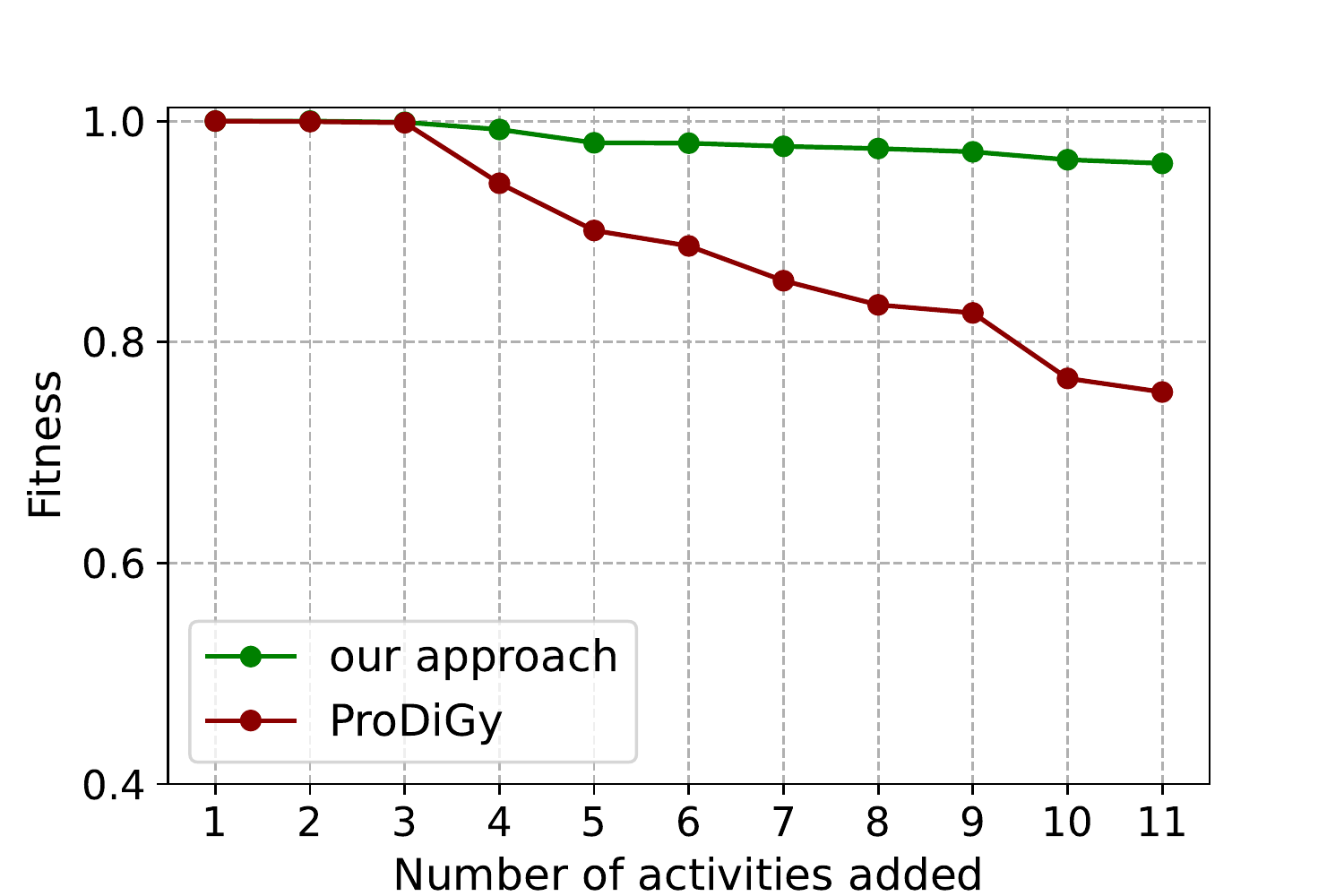}
            \caption{Fitness}\label{fig:exp-patterns-fitness}
        \end{subfigure}
        \begin{subfigure}[b]{0.48\linewidth}
            \includegraphics[width=\linewidth]{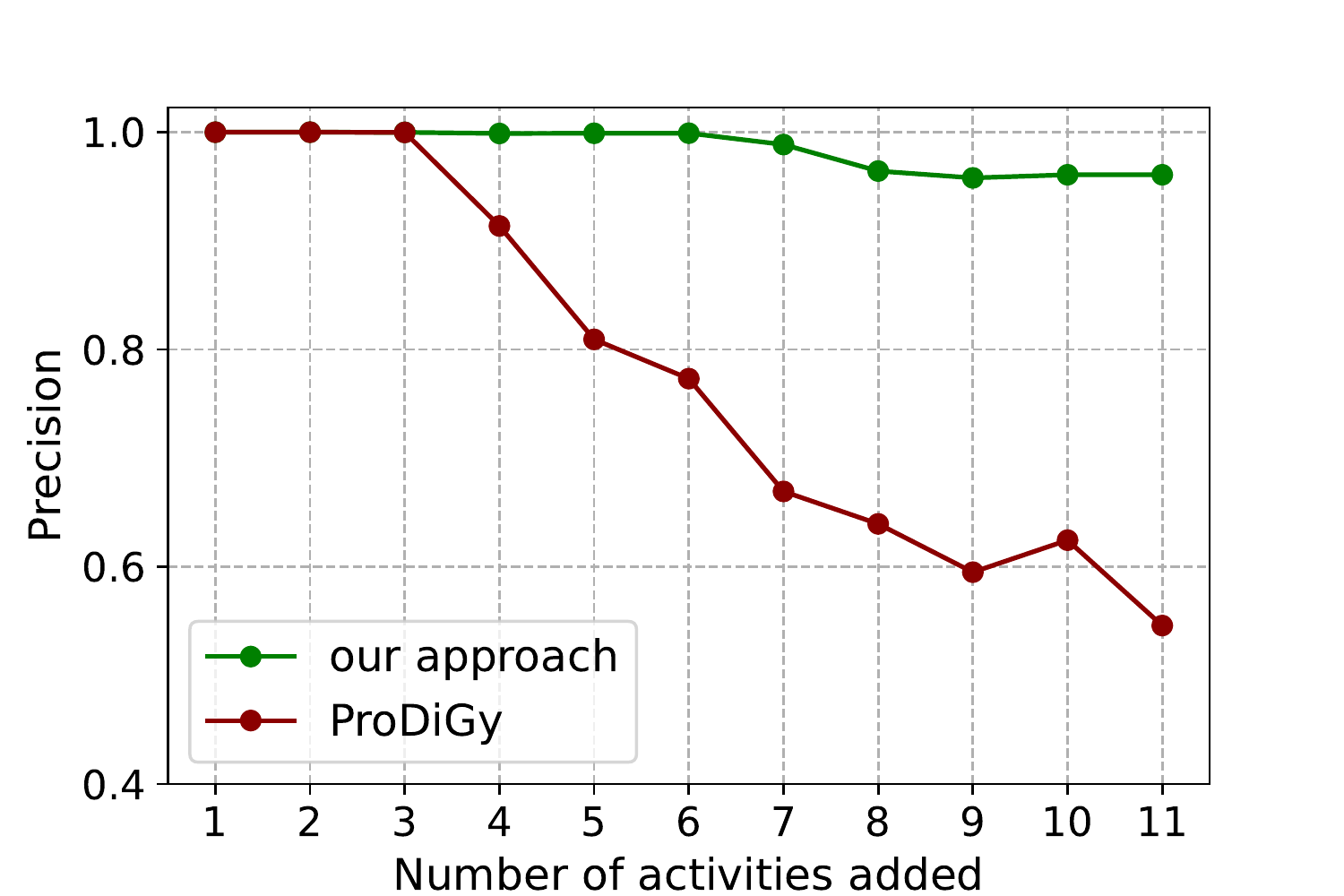}
            \caption{Precision}\label{fig:exp-patterns-precision}
        \end{subfigure}
    \caption{Results on fitness and precision comparison for the effectiveness of patterns}
    \label{fig:exp-patterns-results}
    \vspace{-2em}
\end{figure}

Fig.~\ref{fig:exp-patterns-results} shows the result of the comparison.
The fitness and precision are the average values of the five event logs.
As one can see from the figures, both approaches can capture the behaviors quite well for the first three activities added.
When adding more activities to the model, our approach has consistently higher values for both fitness and precision than ProDiGy.
One might think that this is expected as we extend the set of patterns used in ProDiGy.
However, note that ProDiGy evaluates every possible synthesis rules applications while we only focus on a subset of the nodes using log heuristics.
There is a trade-off between optimal solution and time in our approach.
Nevertheless, the results show that the extended patterns enable us to discover models with higher quality compared to the existing approach, ProDiGy, while limiting the search space.

\vspace{-1em}
\subsubsection{Effects of Ordering Strategy and Search Space Pruning}
\begin{wrapfigure}[13]{r}{.5\textwidth}
  \centering
  \vspace{-3em}
  \includegraphics[width=\linewidth]{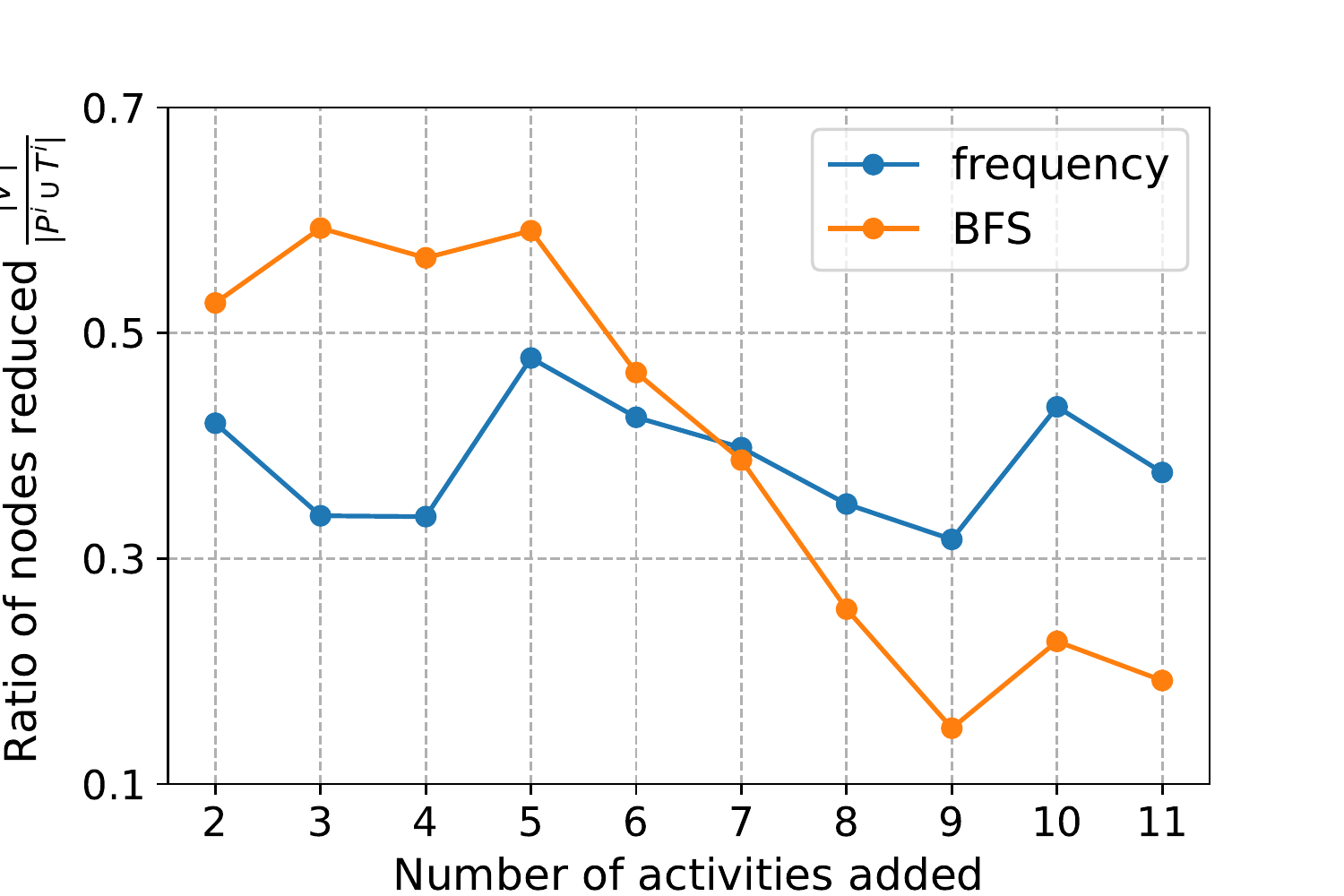}
  \caption{Comparison of the to-be-considered nodes ratio for each iteration between the two ordering strategies.}\label{fig:exp-ordering-strategy}
\end{wrapfigure}

Tab.~\ref{tab:results} shows the results of experiments 2 and 3.
We observe that the BFS-based ordering strategy performs better than the frequency-based strategy (in terms of F1 score and time) for four of the five logs.
We further investigate the reason for the shorter execution time of BFS-based ordering.
As shown in Fig.~\ref{fig:exp-ordering-strategy}, it turns out that the BFS-ordering strategy is more effective (lower $\frac{|V^i|}{|P^i\cup T^i|}$) in reducing the search space at the later stage of the discovery process.
As the model grows, checking the preconditions of an application for the linear dependent place or transition rule becomes more expensive.
Reducing the search space more effectively at the later stage is more beneficial in terms of execution time in most cases.
BFS-based ordering achieves this by considering the closeness of activities in the process.
In such case, activities that are closer together are added first and it is more likely for BFS-based ordering to focus on a smaller subset of nodes on the existing net when pruning the search space compared to the frequency-based one.

\begin{table}[h!]
\vspace{-1em}
\centering
\caption{Results about effects of ordering strategy and comparison to IMf}
\label{tab:results}
\resizebox{0.68\textwidth}{!}{%
\begin{tabular}{@{}cccccccc@{}}
\toprule
Log &
  Miner &
  \begin{tabular}[c]{@{}c@{}}Ordering \\ Strategy\end{tabular} &
  \begin{tabular}[c]{@{}c@{}}IMf\\ filter\end{tabular} &
  Fitness &
  Precision &
  F1 &
  Time (s) \\ \midrule
\multirow{3}{*}{BPI2017A} & ours & frequency             & -   & 0.970 & 0.947 & 0.958 & 734 \\
                          & ours & BFS                   & -   & 0.989 & 0.935 & 0.961 & 342 \\
                          & IMf  & -                     & 0.2 & 0.999 & 0.936 & 0.967 & 10  \\ \midrule
\multirow{3}{*}{BPI2017O} & ours & frequency             & -   & 0.994 & 0.962 & 0.978 & 560 \\
                          & ours & BFS                   & -   & 0.989 & 1.000 & 0.994 & 240 \\
                          & IMf  & -                     & 0.2 & 0.997 & 0.907 & 0.950 & 7   \\ \midrule
\multirow{3}{*}{helpdesk} & ours & frequency             & -   & 0.972 & 0.984 & 0.977 & 54  \\
                          & ours & BFS                   & -   & 0.981 & 0.976 & 0.978 & 44  \\
                          & IMf  & -                     & 0.2 & 0.967 & 0.950 & 0.958 & 1   \\ \midrule
\multirow{3}{*}{\begin{tabular}[c]{@{}c@{}}hospital\\ billing\end{tabular}} &
  ours &
  frequency &
  - &
  0.961 &
  0.810 &
  0.879 &
  567 \\
                          & ours & BFS                   & -   & 0.989 & 0.935 & 0.961 & 407 \\
                          & IMf  & -                     & 0.2 & 0.982 & 0.906 & 0.943 & 45  \\ \midrule
\multirow{3}{*}{traffic}  & ours & frequency             & -   & 0.960 & 0.930 & 0.945 & 321 \\
                          & ours & BFS                   & -   & 0.964 & 0.720 & 0.825 & 427 \\
                          & IMf  & \multicolumn{1}{l}{-} & 0.4 & 0.904 & 0.720 & 0.801 & 28  \\ \bottomrule
\end{tabular}%
}
\vspace{-1em}
\end{table}

\vspace{-1em}
\subsubsection{Effects of Non-block Structures}
\begin{figure}[h!]
    \vspace{-1.5em}
    \centering
        \begin{subfigure}[b]{\linewidth}
            \centering
            \includegraphics[width=0.87\linewidth]{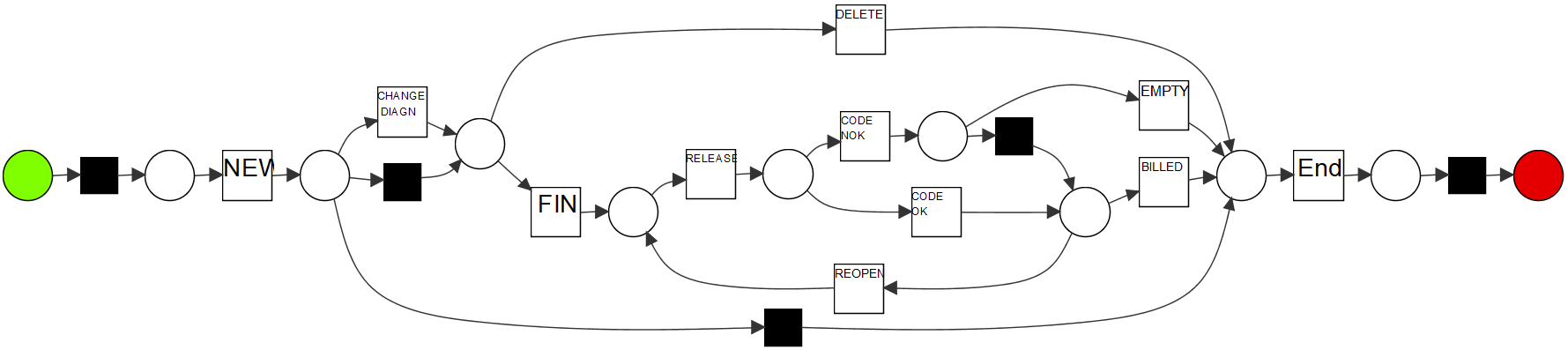}
            \caption{The discovered model using our approach. Due to the more flexible structure, one can execute \textit{EMPTY}, \textit{BILLED}, or \textit{REOPEN} after \textit{CODE NOK} while only \textit{BILLED} or \textit{REOPEN} are executable after \textit{CODE OK}. The construct is not discoverable by IMf. }\label{fig:result-model-ours}
        \end{subfigure}
        \begin{subfigure}[b]{\linewidth}
            \centering
            \includegraphics[width=0.87\linewidth]{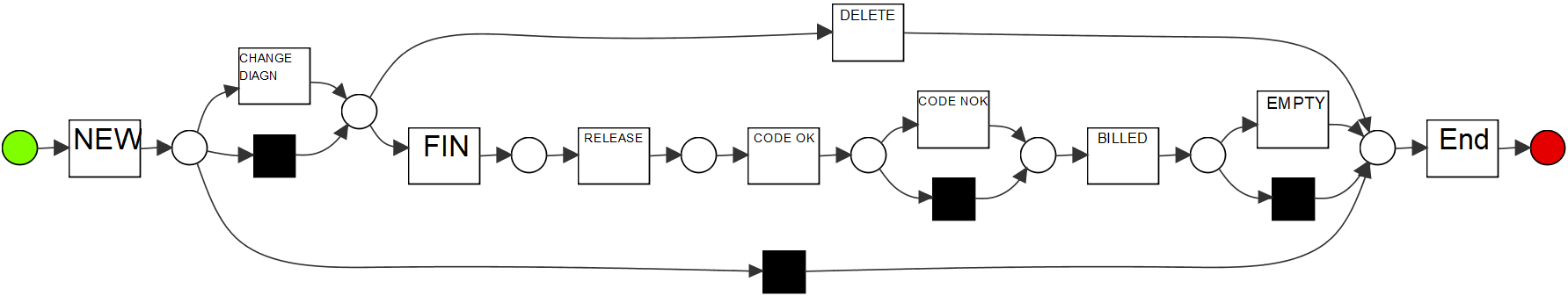}
            \caption{The discovered model using IMf. Note that activity \textit{REOPEN} is dropped by the filter of IMf.}\label{fig:result-model-imf}
        \end{subfigure}
    \caption{The models discovered by our approach and IMf for the hospitalBilling log.}
    \label{fig:resulting-models}
    \vspace{-1.5em}
\end{figure}

Table~\ref{tab:results} shows that compared to IMf, the models discovered by our approach have higher F1 scores for four of the five logs.
Note that the fitness values of the models discovered by our approach are all higher than the defined threshold $0.95$.
In general, IMf tends to discover models with higher fitness values while our approach discovers models with higher precision.
In IMf, one can use the filter threshold to balance fitness and precision. 
This is also the case in our approach, the user can set a lower fitness threshold to include more candidate nets that are less fitting but more precise.
Fig.~\ref{fig:resulting-models} shows the discovered models from the two approaches for the hospitalBilling log.
While the overall structure of Fig.~\ref{fig:result-model-ours} is similar to its counterpart in Fig.~\ref{fig:result-model-imf}, our approach discovered non-block structures at the later stage of the process.
Such construct is not possible to model by IMf.
The result shows that our approach can discover sound free-choice workflow nets with non-block structures and produce competitive model quality as the state-of-the-art algorithm.

\vspace{-1em}
\section{Conclusion and Future Work}\label{sec:conclusion}
In this paper, we present a discovery algorithm that aims to discover sound free-choice workflow nets with non-block structures.
The algorithm utilizes the synthesis rules to incrementally add activities with predefined patterns to discover models that are guaranteed to be sound and free-choice.
Moreover, a certain level of replay fitness is guaranteed by a user-defined threshold.

The approach has been implemented and evaluated using various real-life event logs. 
The results show that the process models discovered by our approach have higher model quality (in terms of both replay fitness and precision) than the existing approach~\cite{DixitBA18ProDiGy}, which also depends on synthesis rules.
Moreover, our approach produces competitive model quality compared to the state-of-the-art: Inductive Miner - infrequent.
For future work, we plan to explore more advanced ordering strategies and investigate their influences on the model quality and computation time.
The other direction is to further speed up the approach as the long execution time is a clear limitation.
This could be done by exploiting the log-based heuristics further.

\subsubsection{Acknowledgements.}
We thank the Alexander von Humboldt (AvH) Stiftung for
supporting our research.

\bibliographystyle{splncs04}

\typeout{}
\bibliography{myrefs}

\end{document}